\documentclass[prd,twocolumn,showpacs,aps,epsbox]{revtex4}
\usepackage{epsfig}

\def\beq{\begin{equation}}
\def\eeq{\end{equation}}
\def\bea{\begin{eqnarray}}
\def\eea{\end{eqnarray}}
\def\nn{\nonumber\\}
\def\pa{\partial}
\def\ra{\rightarrow}
\def\bp{\mbox{\boldmath$\phi$}}
\def\tm{\tilde{m}}
\def\te{\tilde{e}}
\def\tq{\tilde{q}}
\def\tE{\tilde{E}}
\def\tQ{\tilde{Q}}
\def\tp{\tilde{\phi}}

\begin{document}

\draft
\title{Unified pictures of Q-balls and Q-tubes}
\author{Takashi Tamaki}
\email{tamaki@ge.ce.nihon-u.ac.jp}
\affiliation{Department of Physics, General Education, College of Engineering, 
Nihon University, Tokusada, Tamura, Koriyama, Fukushima 963-8642, Japan}
\author{Nobuyuki Sakai}
\email{nsakai@e.yamagata-u.ac.jp}
\affiliation{Department of Education, Yamagata University, Yamagata 990-8560, Japan}

\begin{abstract}
While Q-balls have been investigated intensively for many years, another type of nontopological 
solutions, Q-tubes, have not been understood very well.
In this paper we make a comparative study of Q-balls and Q-tubes.
First, we investigate their equilibrium solutions for four types of potentials.
We find, for example, that in some models the charge-energy relation is similar between Q-balls and Q-tubes while in other models the relation is quite different between them.
To understand what determines the charge-energy relation, which is a key of stability of the equilibrium solutions, we establish an analytical method to obtain the two limit values of the energy and the charge.
Our prescription indicates how the existent domain of solutions and their stability depends on their shape as well as potentials, which would also be useful for a future study of Q-objects in higher-dimensional spacetime. 
\end{abstract}

\maketitle

\section{Introduction}
Among nontopological solitons, Q-balls have attracted much attention because
they can exist in all supersymmetric extensions of the Standard Model \cite{Kus97b-98}.
Specifically, they can be produced efficiently in the Affleck-Dine (AD) mechanism \cite{AD} and could be responsible for baryon asymmetry \cite{SUSY} and dark matter \cite{SUSY-DM}.
Q-balls can also influence the fate of neutron stars \cite{Kus98}.
Based on these motivations, stability of Q-balls has been intensively 
studied \cite{stability,PCS01,SS,TS}.

In spite of these concerns about Q-balls, other equilibrium solutions have not been studied so much, 
while topological defects have several types according to the symmetry. 
For example, observational consequences by cosmic strings, such as gravitational lenses and the 
gravitational wave have been argued for years~\cite{strings}. 

From this point of view, other types of nontopological solutions may play an important role in the the universe.
Recently, two types of nontopological solutions was discussed: Q-tubes and Q-crust, which mean 
tube-shaped (or string-like) and crust-shaped solutions, respectively \cite{SIN}.
As for Q-tubes, some numerical studies manifested sign of their appearance.
First, it has been reported that a filament structure appears just before Q-ball formation in the
numerical simulations~\cite{EJ01}. Second, according to the simulations of the collision of 
two Q-balls, two apparent rings are formed \cite{Tsuma}.
We conjecture that the filament structure and the rings are Q-tubes.

In \cite{SIN} numerical solutions were investigated for the potential,
\beq\label{V3}
V_{3}(\phi):={m^2\over2}\phi^2-\mu\phi^3+\lambda\phi^4
~~{\rm with}~~
m^2,~\mu,~\lambda>0, 
\eeq
which we call the $V_{3}$ model.
In the case of Q-balls \cite{stability,PCS01,SS}, however, the charge-energy relation, which is a key of stability of the equilibrium solutions, is quite dependent on potentials $V(\phi)$.
Therefore, our first concern is how Q-tube solutions depend on potentials.

Our second concern is how different in the charge-energy relation between Q-tubes and Q-balls.
This shape-dependence is closely related to the dimension-dependence because a cylindrical Q-tube in 3+1 spacetime is equivalent to a ``Q-ball" in 2+1 spacetime if we ignore gravity.
If this dimension-dependence becomes manifest, it would be useful for investigating 
other Q-objects or those in higher-dimensional spacetime \cite{TCS}.

For these reasons, in this paper, we make a comparative study of Q-balls and Q-tubes.
This paper is organized as follows.
In Sec. II, we explain briefly what Q-balls and Q-tubes are.
In Sec. III, we investigate their equilibrium solutions numerically for four types of potentials.
In Sec. IV, we evaluate analytically the limit values of the energy and the charge.
In Sec. V, we devote to concluding remarks.

\section{Equilibrium solutions}

Consider an SO(2)-symmetric scalar field $\bp=(\phi_1,\phi_2)$, whose action is given by
\beq\label{S}
{\cal S}=\int d^4x\left[
-\frac12\eta^{\mu\nu}\pa_{\mu}\bp\cdot\pa_{\nu}\bp-V(\phi) \right],
~~\phi\equiv\sqrt{\bp\cdot\bp}.
\eeq

\subsection{Q-balls}

For a Q-ball, we assume spherical symmetry and homogeneous phase rotation,
\beq\label{qball-phase}
\bp=\phi(r)(\cos\omega t,\sin\omega t).
\eeq
One has a field equation,
\beq\label{FEqball}
{d^2\phi\over dr^2}+\frac2r{d\phi\over dr}+\omega^2\phi={dV\over d\phi}.
\eeq
This is equivalent to the field equation for a single static scalar field with an effective potential
\beq
V_{\omega}=V-\frac12\omega^2\phi^2.
\eeq

Equilibrium solutions $\phi(r)$ with a boundary condition
\beq\label{BCqball}
{d\phi\over dr}(r=0)=0,~~~\phi(r\ra\infty)=0,
\eeq
exist if min$(V_{\omega}) <V_{\omega}(0)$ and $d^2V_{\omega}/d\phi^2(0) > 0$. 
This condition is rewritten as
\beq\label{Qexist}
{\rm min}\left[{2V\over\phi^2}\right]<\omega^2<m^2\equiv{d^2V\over d\phi^2}(0),
\eeq
where we have put $V(0)=0$ without loss of generality.

For a Q-ball solution, we can define 
the energy and the charge, respectively, as
\bea\label{EQdef}
E&=&4\pi\int_0^{\infty}r^2 dr
\left\{{1\over2}\omega^2\phi^2+\frac12\left({d\phi\over dr}\right)^2+V\right\},\nn
Q&=&4\pi\omega\int_0^{\infty}r^2\phi^2dr. \label{EQball-definition}
\eea
The $Q$-$E$ relation is a key to understand stability of equilibrium solutions in terms of catastrophe theory \cite{SS}.

\begin{figure}[htbp]
\psfig{file=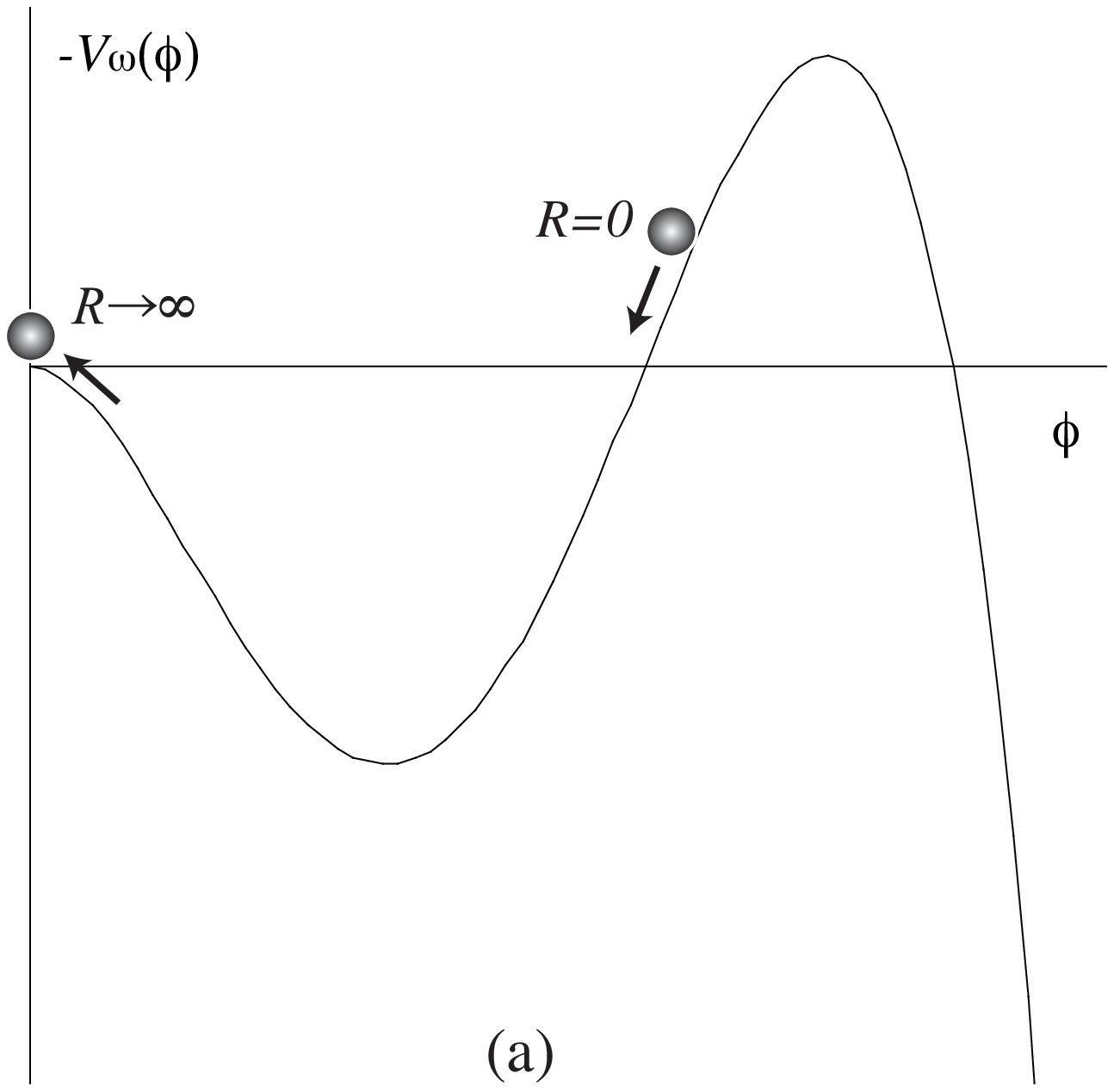,width=3in}
(a)
\psfig{file=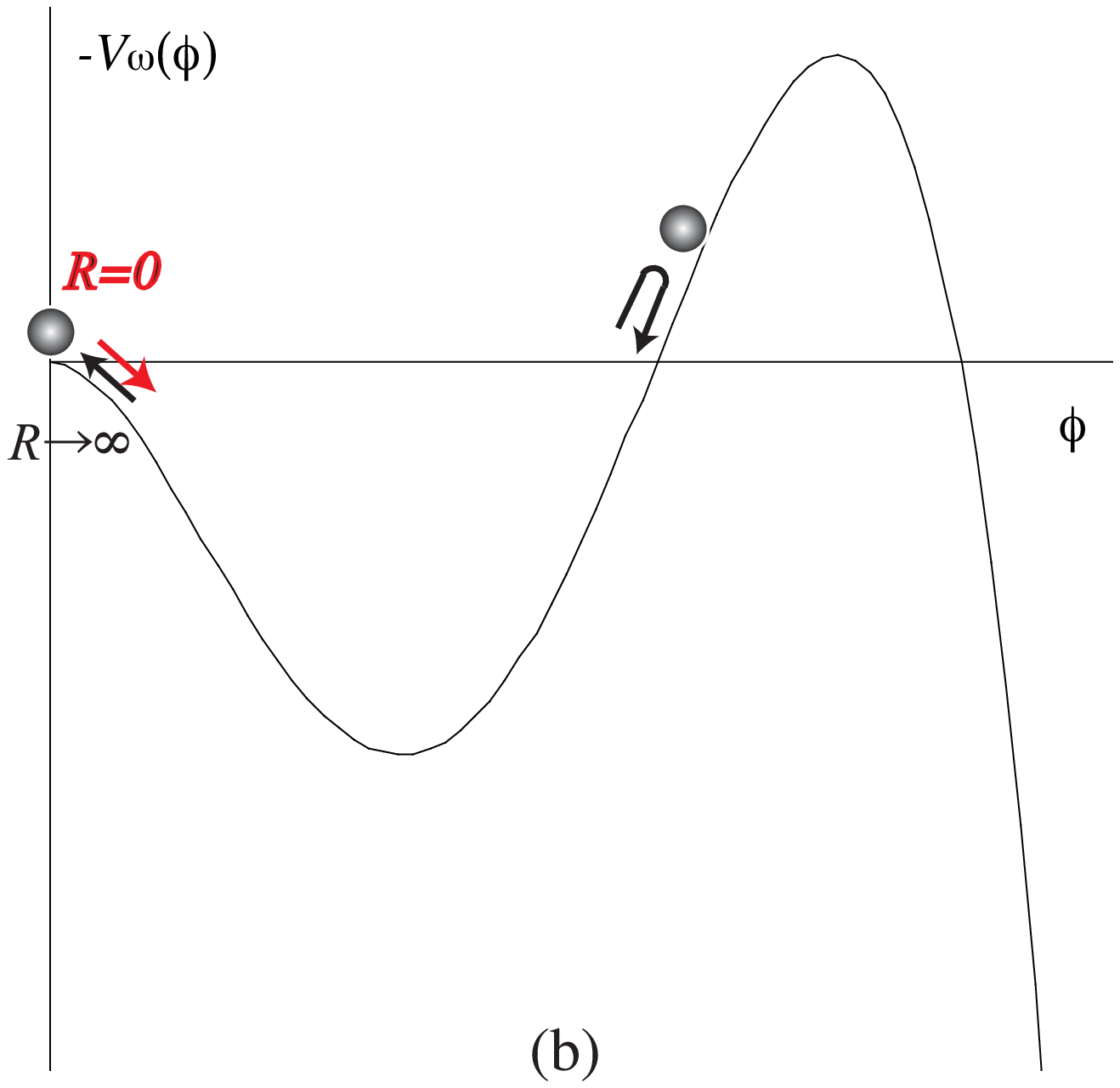,width=3in}
(b)
\caption{\label{Newton}
Interpretation of (a) Q-balls and $n=0$ solutions in Q-tubes and (b) $n\geq 1$ solutions 
in Q-tubes by analogy with a particle motion in Newtonian mechanics.}
\end{figure}

\subsection{Q-tubes}

For a Q-tube, we suppose a string-like configuration,
\beq\label{qtube-phase}
\bp=\phi(R)(\cos(n\varphi+\omega t),\sin(n\varphi+\omega t)),
\eeq
where $n$ is nonnegative integer and $(R,\varphi,z)$ is the cylindrical coordinate system.
The field equation becomes
\beq\label{FEtube}
{d^2\phi\over dR^2}+\frac1R{d\phi\over dR}-{n^2\phi\over R^2}+\omega^2\phi={dV\over d\phi}.
\eeq
In the case of $n=0$, the field equation is the same as (\ref{FEqball}) except for a numerical coefficient.
Therefore, Q-ball-like solutions of $\phi(R)$ exist if the condition (\ref{Qexist}) is satisfied.

In the case of $n\ge1$,  there is no regular solution which satisfies $\phi(0)\ne0$.
However, if we adopt a different boundary condition, 
\beq
\phi(R=0)=\phi(R\ra\infty)=0,
\eeq
there is a new type of regular solutions.
We introduce an auxiliary variable $\psi$ which is defined by
$\phi(R)=R^n\psi(R)$,
Then, Eq.(\ref{FEtube}) becomes
\beq
{d^2\psi\over dR^2}+{2n+1\over R}{d\psi\over dR}+\omega^2\psi
=R^{-n}{dV\over d\phi}\Big|_{\phi=R^n\psi}
\eeq
If we choose $\psi(0)$ appropriately, we obtain a solution $\psi(R)$ which is expressed in the Maclaurin series without odd powers in the neighborhood of $R=0$.
In terms of the original variable $\phi(R)$, the $n$th differential coefficient $\phi^{(n)}(0)=\psi(0)$ should be determined by the shooting method, while any lower derivative vanishes at $R=0$.

In the same way as for Q-balls \cite{Col85},
existence of Q-tube solutions can be interpreted as follows.
If one regards the radius $R$ as \lq time\rq\ and the scalar amplitude 
$\phi(R)$ as \lq the position of
a particle\rq, one can understand $n=0$ solutions in words of 
Newtonian mechanics, as shown in Fig.\ \ref{Newton}(a).
Equation (\ref{FEtube}) describes a one-dimensional motion of a particle 
under the conserved force due to the potential $-V_{\omega}(\phi)$ and 
the \lq time\rq-dependent friction $-(1/R)d\phi/dR$.
If one chooses the \lq initial position\rq\ $\phi(0)$ appropriately, 
the static particle begins to roll down the potential slope, climbs up and approaches 
the origin over infinite time. 

Similarly, we can also understand $n\ge1$ solutions as shown in Fig.\ \ref{Newton}(b).
In this case, there are two non-conserved forces, the friction $-(1/R)d\phi/dR$ and 
the repulsive force $n^2\phi^2/R^2$.
If $n=1$,  by choosing the  \lq initial velocity\rq\ $d\phi/dR(0)$ appropriately, the particle goes down and up the slope, and at some point $\phi=\phi_{{\rm max}}$ it turns back and approaches the origin over infinite time.
If $n\ge2$, $d\phi/dR(0)$ vanishes; instead, the $n$th derivative $\phi^{(n)}(0)$ gently pushes the particle at $\phi=0$. Therefore, with the appropriate choice of $\phi^{(n)}(0)$, the particle moves along a similar trajectory to that of $n=1$.
This argument also indicates that the existence condition of 
$n\geq 1$ solutions are the same as that of $n=0$ solutions, (\ref{Qexist}).
Solutions with the same behavior as the $n=1$ solutions were obtained by Kim {\it et al}.\cite{Kim}, who studied the SO(3)-symmetric scalar field without Q-charge.

Because our Q-ball solutions are infinitely long, the energy and the charge (\ref{EQdef}) diverge.
We therefore define the energy and the charge per unit length, respectively, as
\bea
{e}&=&2\pi\int_0^{\infty}RdR
\left\{{1\over2}\omega^2\phi^2+\frac12\left({d\phi\over dR}\right)^2+\frac{n^2\phi^2}{2R^2}
+V\right\},\nn
{q}&=&2\pi\omega\int_0^{\infty}R\phi^2dR. \label{eq-definition}
\eea

\begin{table}
\begin{tabular}{|c|c|c|}\hline
& lower limit of $\omega^2$ & upper limit of $\omega^2$ \\\hline
Type I: min$[V]=0$ & min$[2V/\phi^2]$ (thin) & $m^2$ (thick) \\\hline
Type II: min$[V]<0$ & 0 & $m^2$ (thick) \\\hline
\end{tabular}
\caption{Two types of Q-balls/Q-tubes solutions and two limits of $\omega^2$.}
\end{table}

\subsection{Two types and two limits}

The existence condition (\ref{EQdef}) indicates that both Q-balls and Q-tubes are classified into two types of solutions, according to the sign of min$[V(\phi)]$. 

{\bf Type I: min$[V(\phi)]=V(0)=0$}. In this case min$[2V/\phi^2]$ is also positive and the lower limit of $\omega$. The two limits $\omega^2\ra$min$[2V/\phi^2]$ and $\omega^2\ra m^2$ correspond to the thin-wall limit and the thick-wall limit, respectively.

{\bf Type II: min$[V(\phi)]<0$}. In this case min$[2V/\phi^2]$ is negative. Because $\omega^2>0$, there is no thin-wall limit, $\omega^2\ra$min$[2V/\phi^2]$. The thick-wall limit, $\omega^2\ra m^2$, still exists.

The two limits of $\omega^2$ for the two types of solutions are summarized in Table I.

\begin{figure}[htbp]
\psfig{file=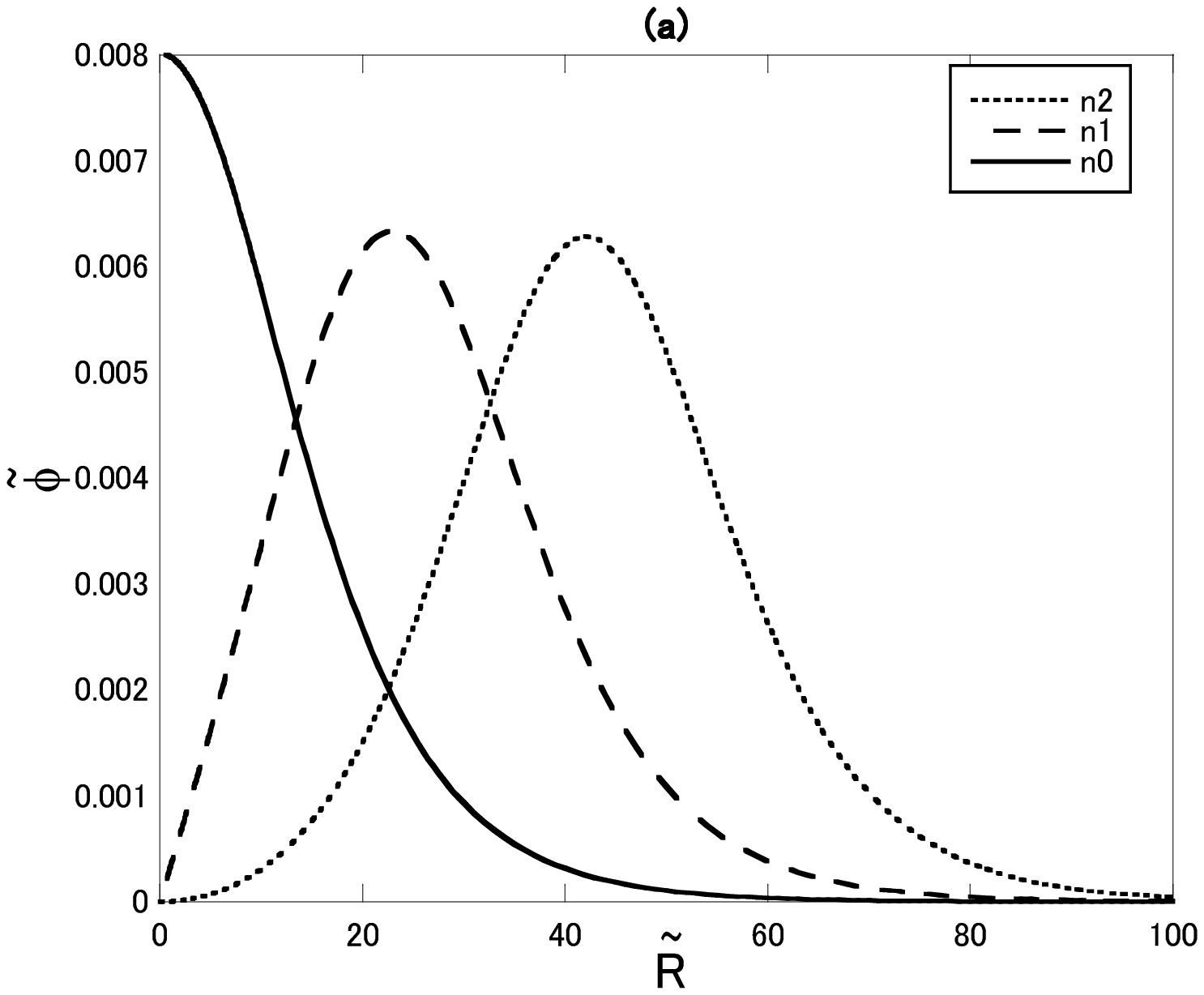,width=3.2in}
\psfig{file=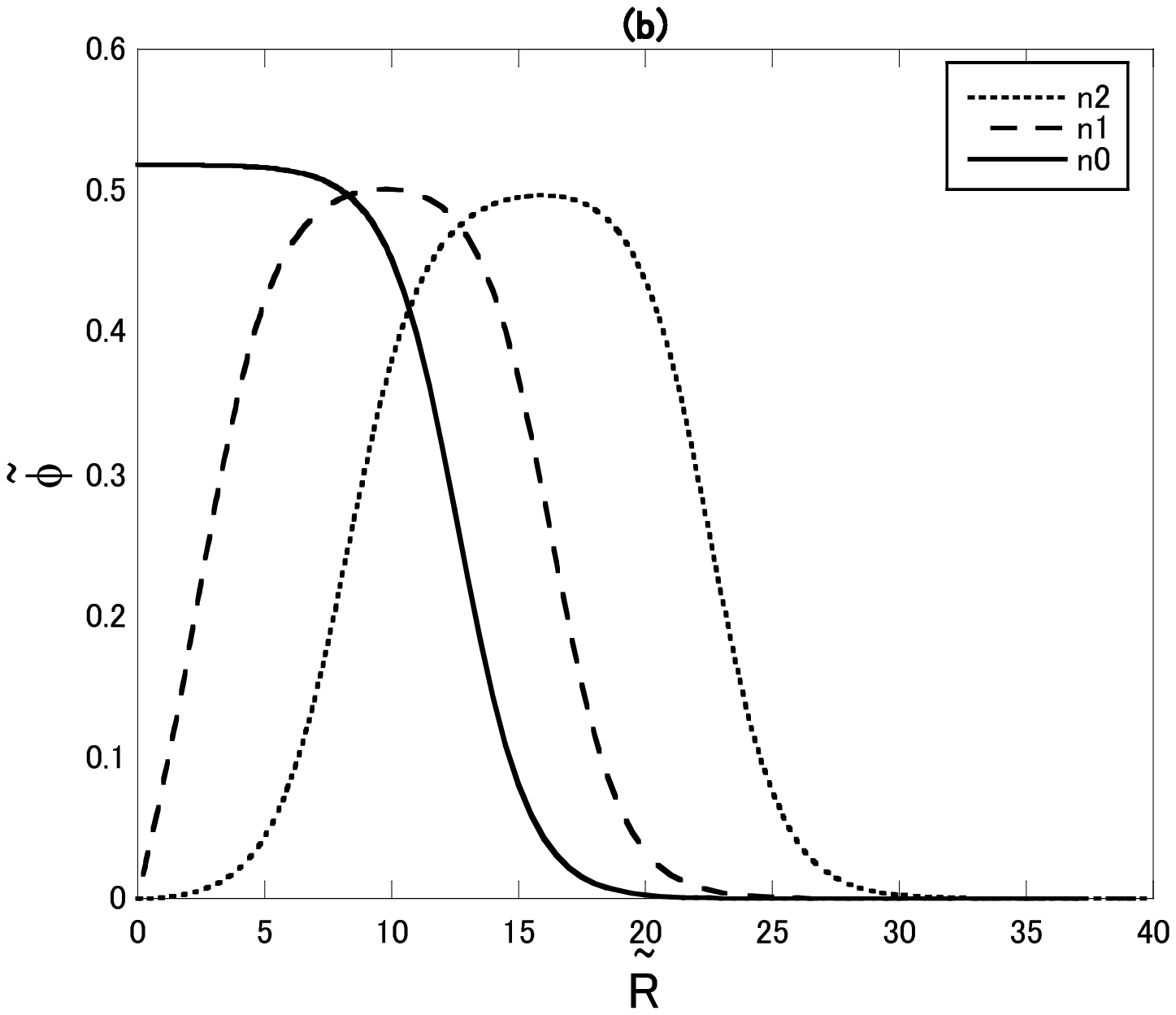,width=3.2in}
\caption{\label{field-V3}
The field configurations of the scalar field for Q-tubes in the $V_{3}$ model with 
$\tm^2 =0.6$ (Type I): (a) $\epsilon^2=0.01$ (thick-wall) and $\epsilon^2=0.48$ (thin-wall).}
\end{figure}
\begin{figure}[htbp]
\psfig{file=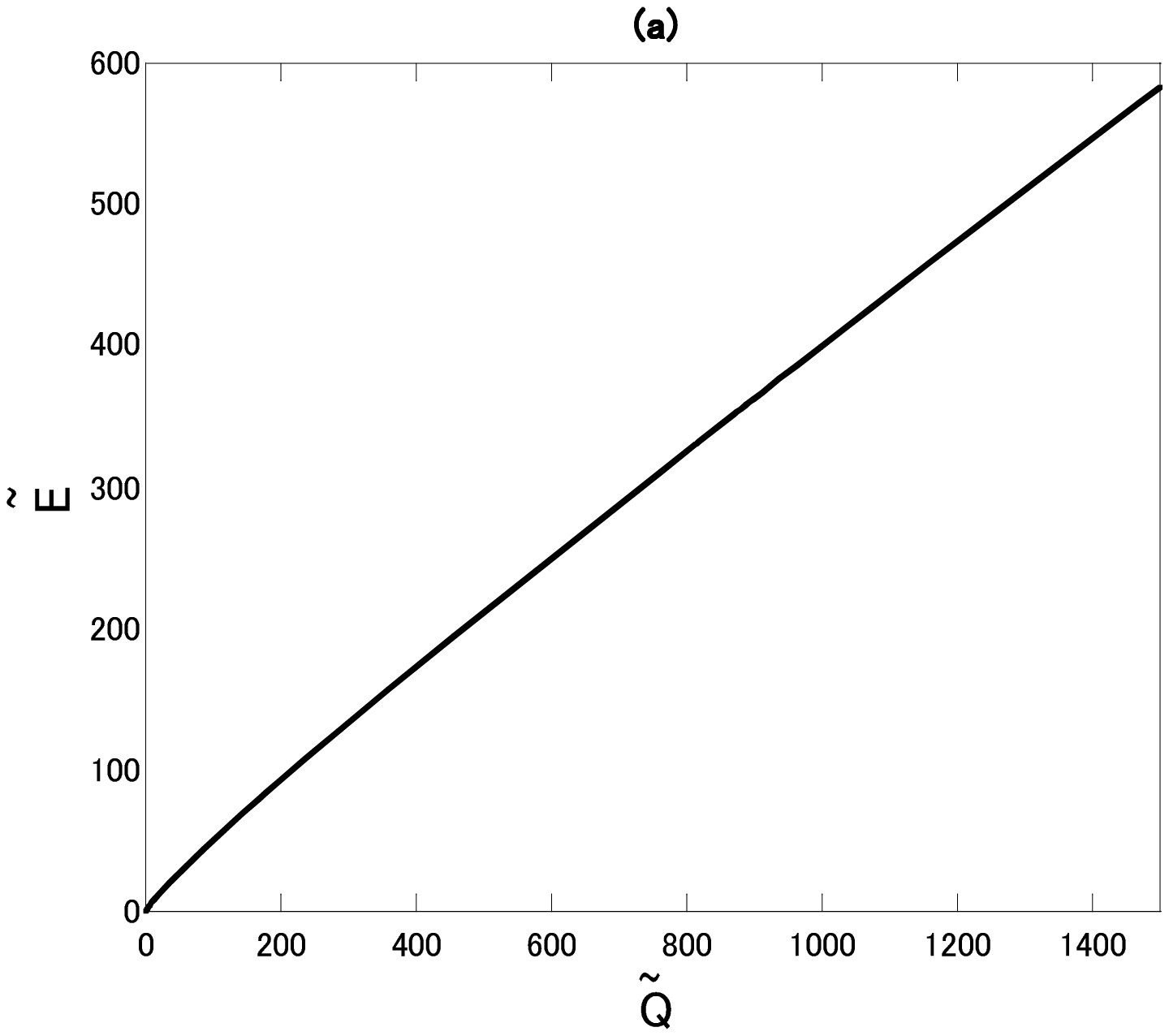,width=3.2in}
\psfig{file=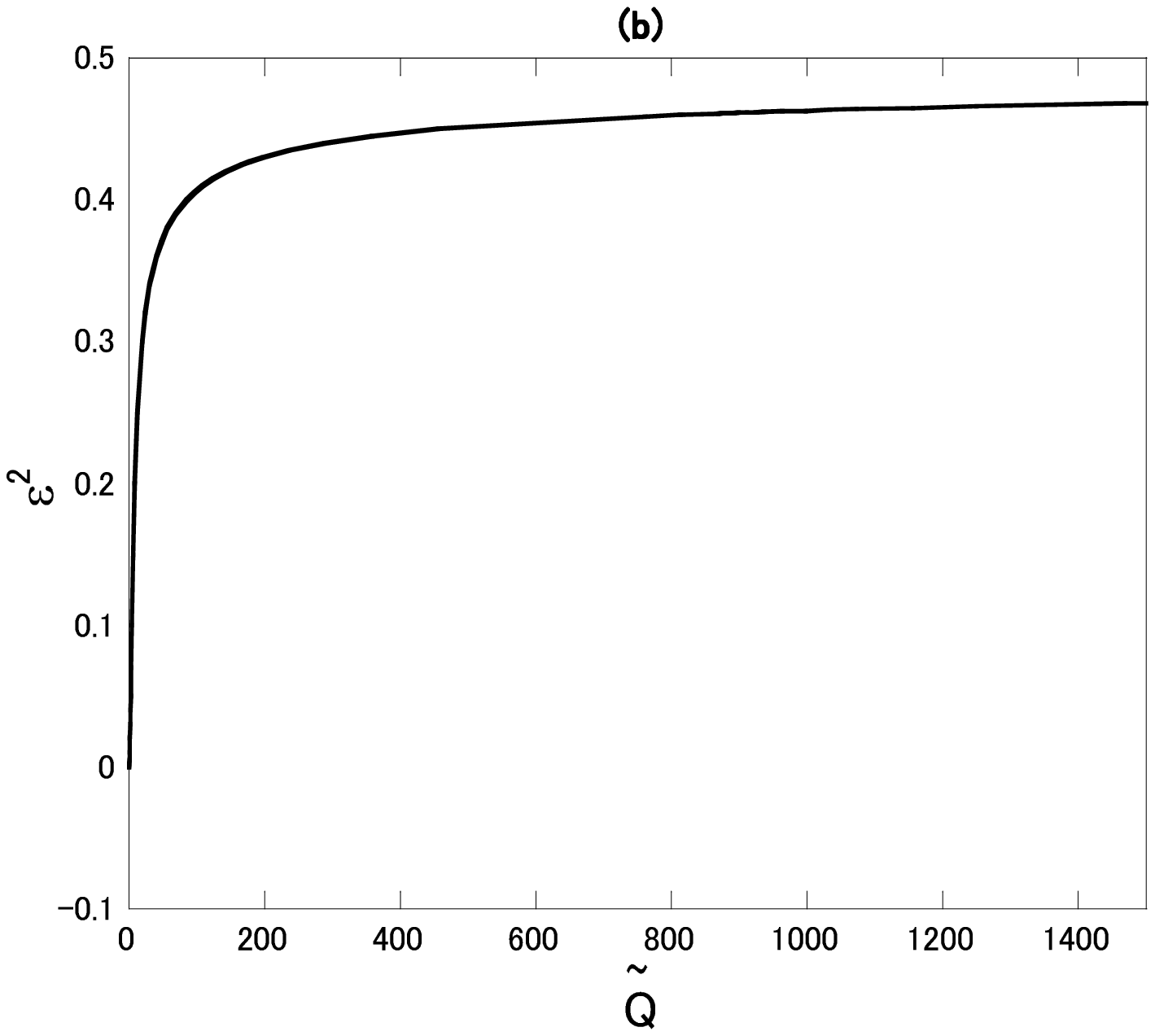,width=3.2in}
\caption{\label{QEQball-V3m06}
(a) $\tilde{Q}$-$\tilde{E}$ and (b) $\tilde{Q}$-$\epsilon^2$ relations for Type I Q-balls in the $V_{3}$ model: $\tilde{m}^{2}=0.6$.}
\end{figure}
\begin{figure}[htbp]
\psfig{file=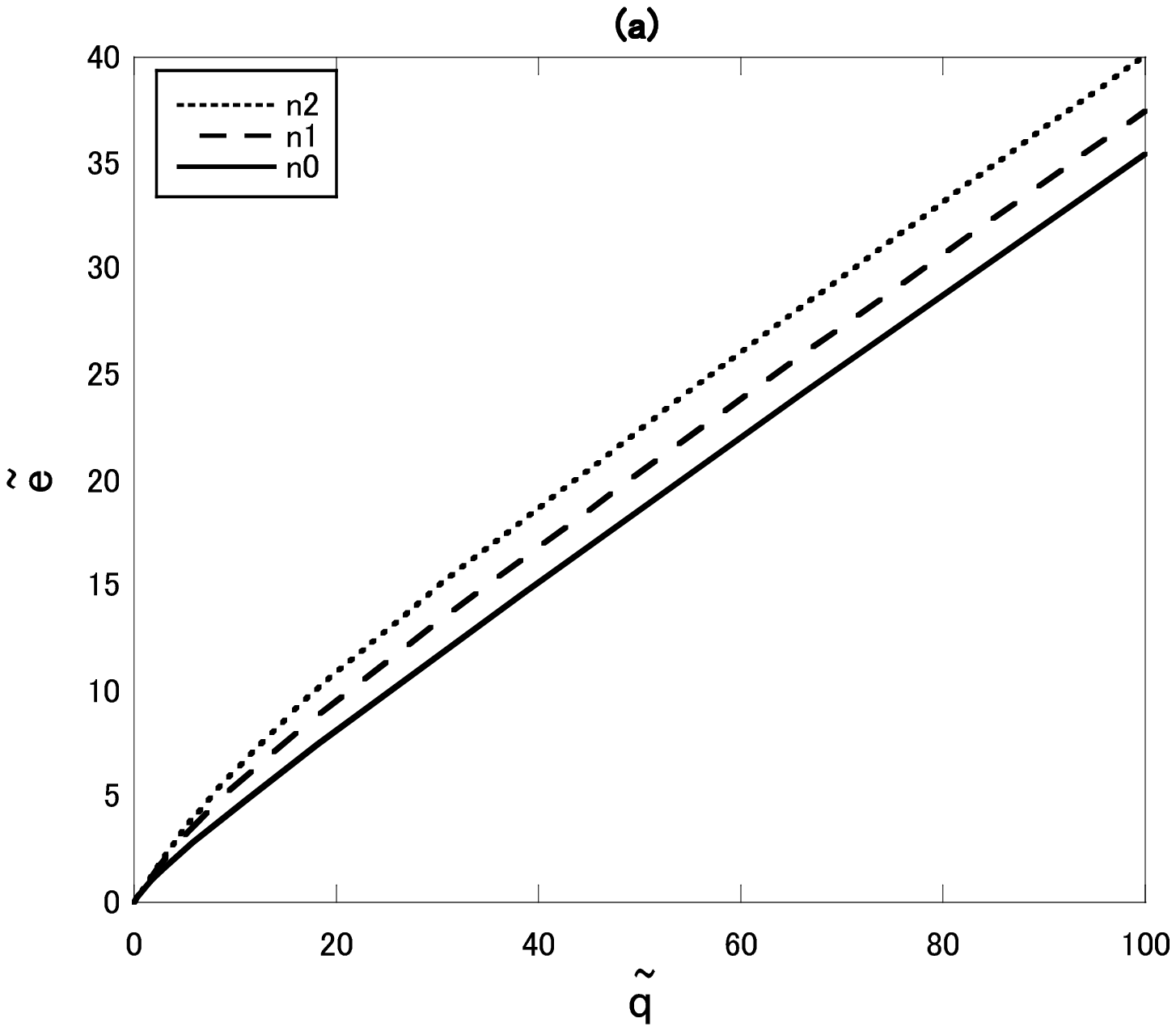,width=3.2in}
\psfig{file=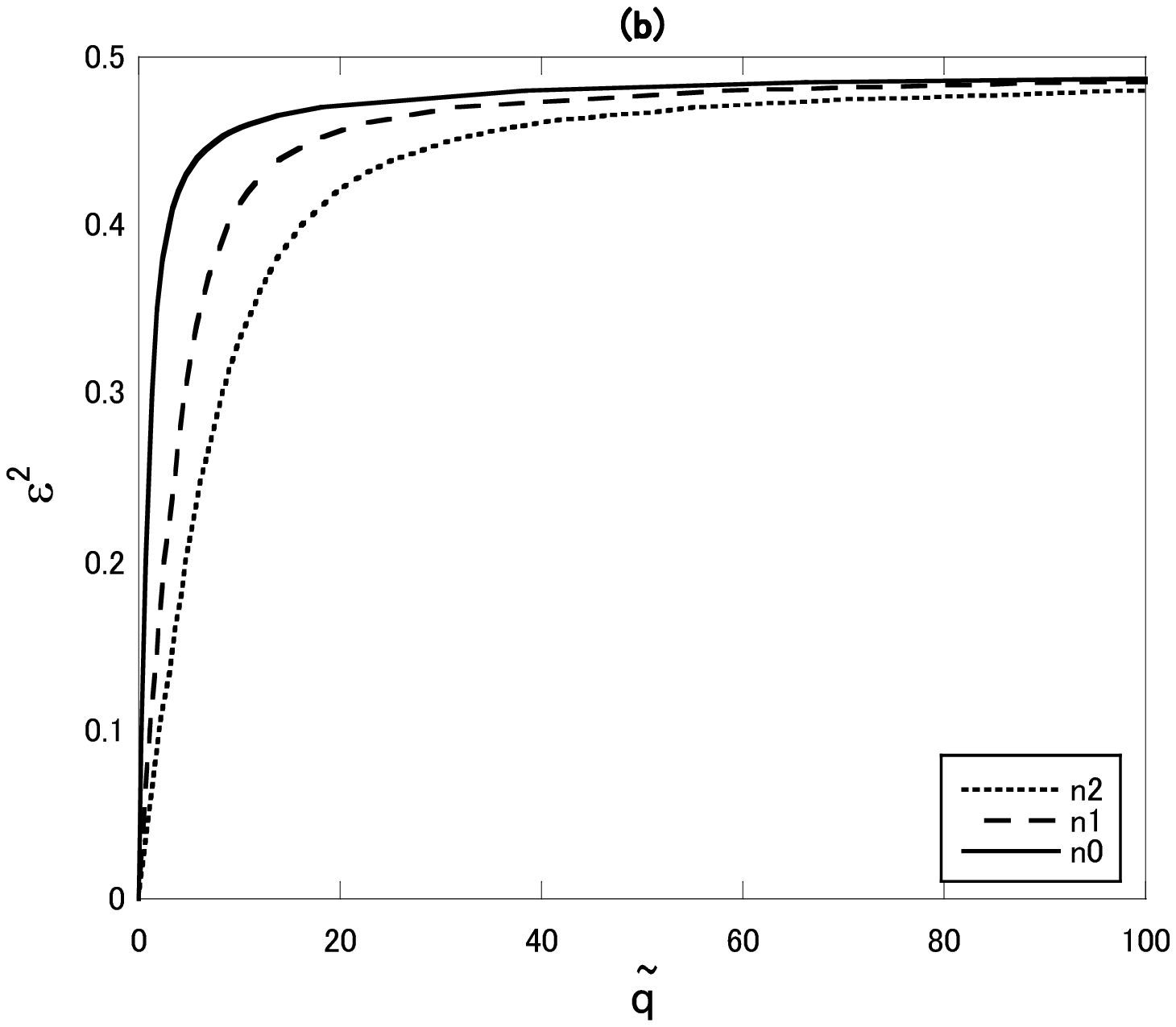,width=3.2in}
\caption{\label{qe-V3m06}
(a) $\tq$-$\te$ and (b) $\tq$-$\epsilon^2$ relations for Type I Q-tubes in the $V_{3}$ model:
$\tilde{m}^{2}=0.6$.}
\end{figure}
\begin{figure}[htbp]
\psfig{file=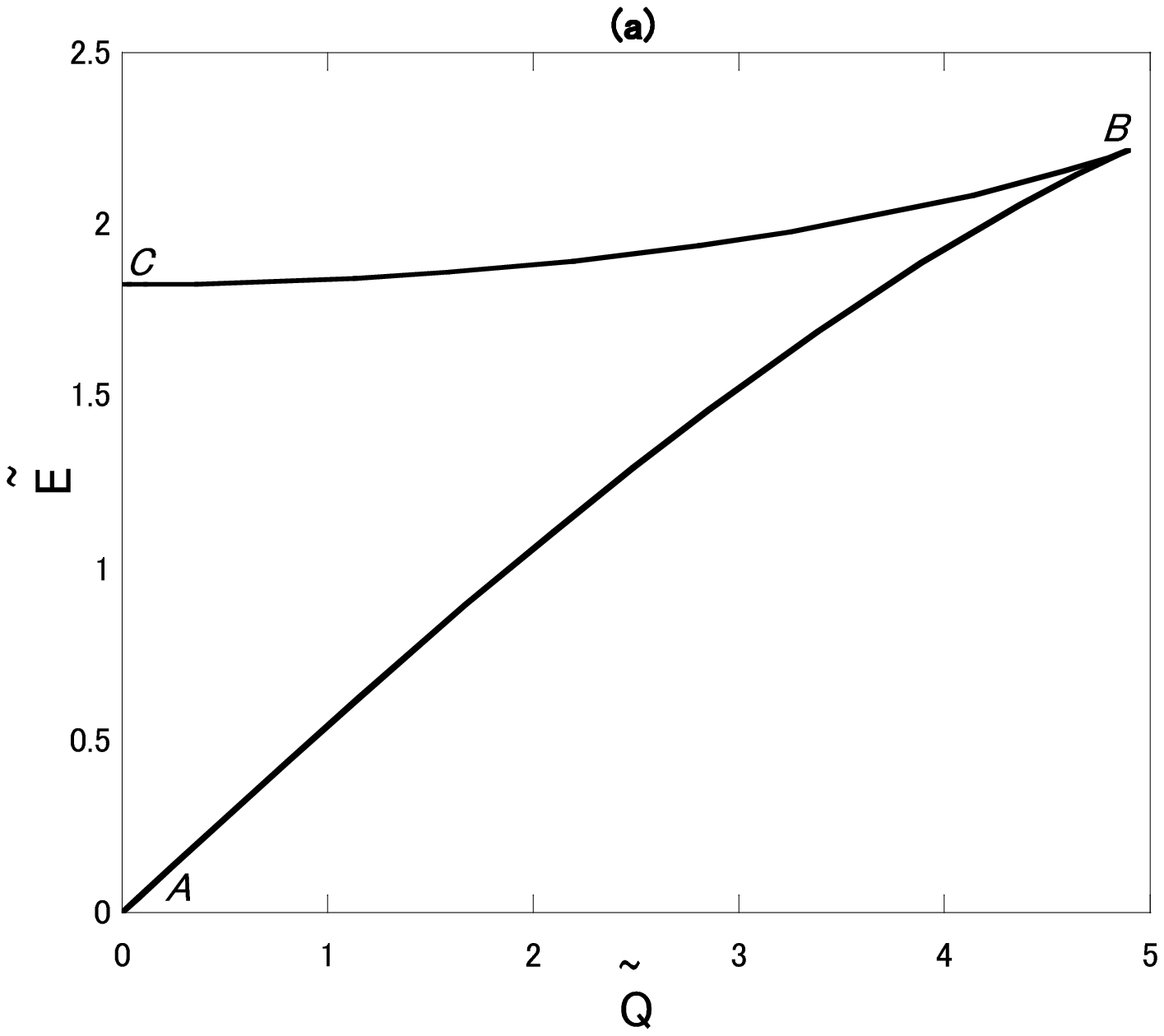,width=3.2in}
\psfig{file=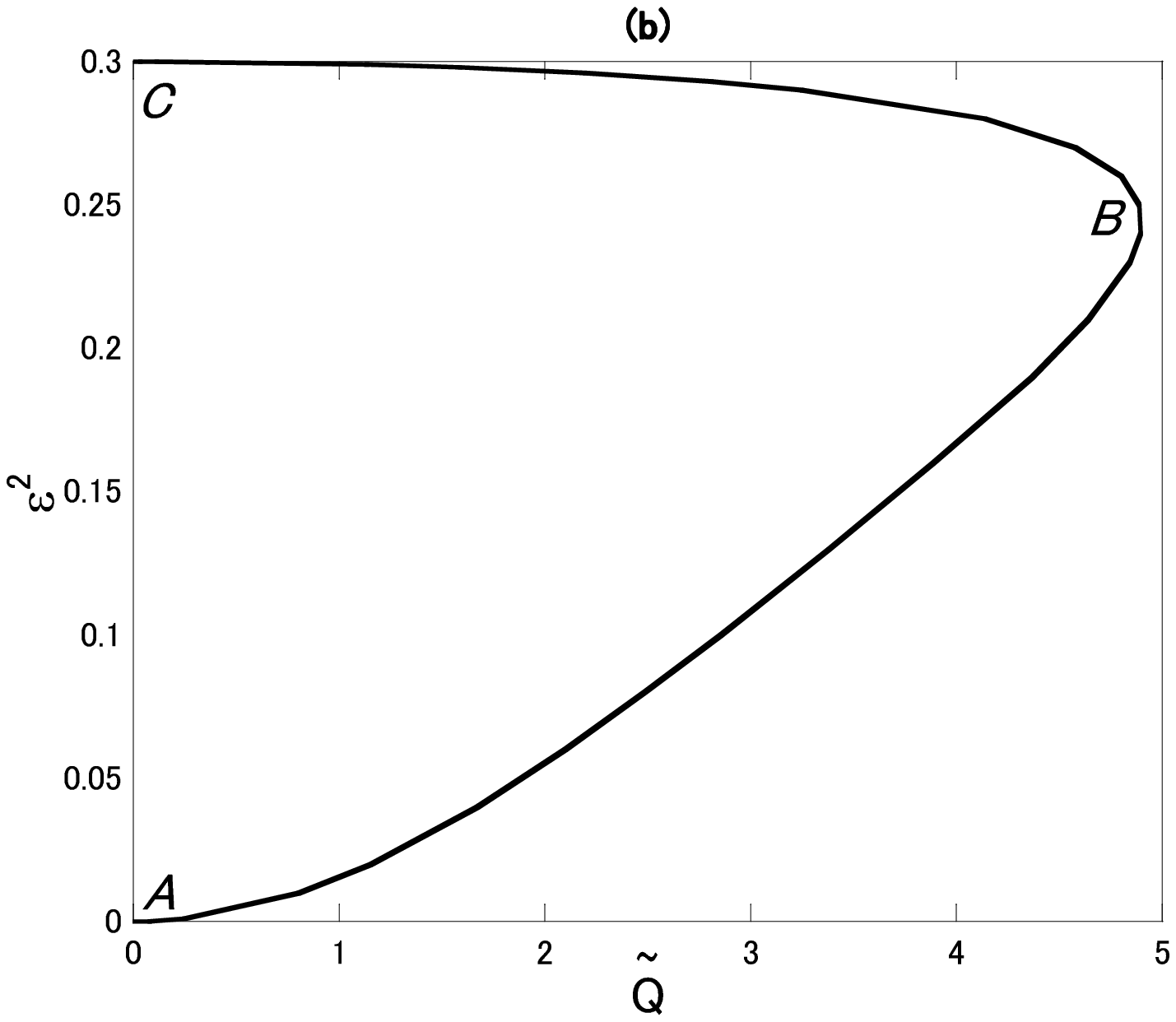,width=3.2in}
\caption{\label{QEQball-V3m03}
(a) $\tilde{Q}$-$\tilde{E}$ and (b) $\tilde{Q}$-$\epsilon^2$ relations for Type I Q-balls in the $V_{3}$ model: $\tilde{m}^{2}=0.3$.}
\end{figure}
\begin{figure}[htbp]
\psfig{file=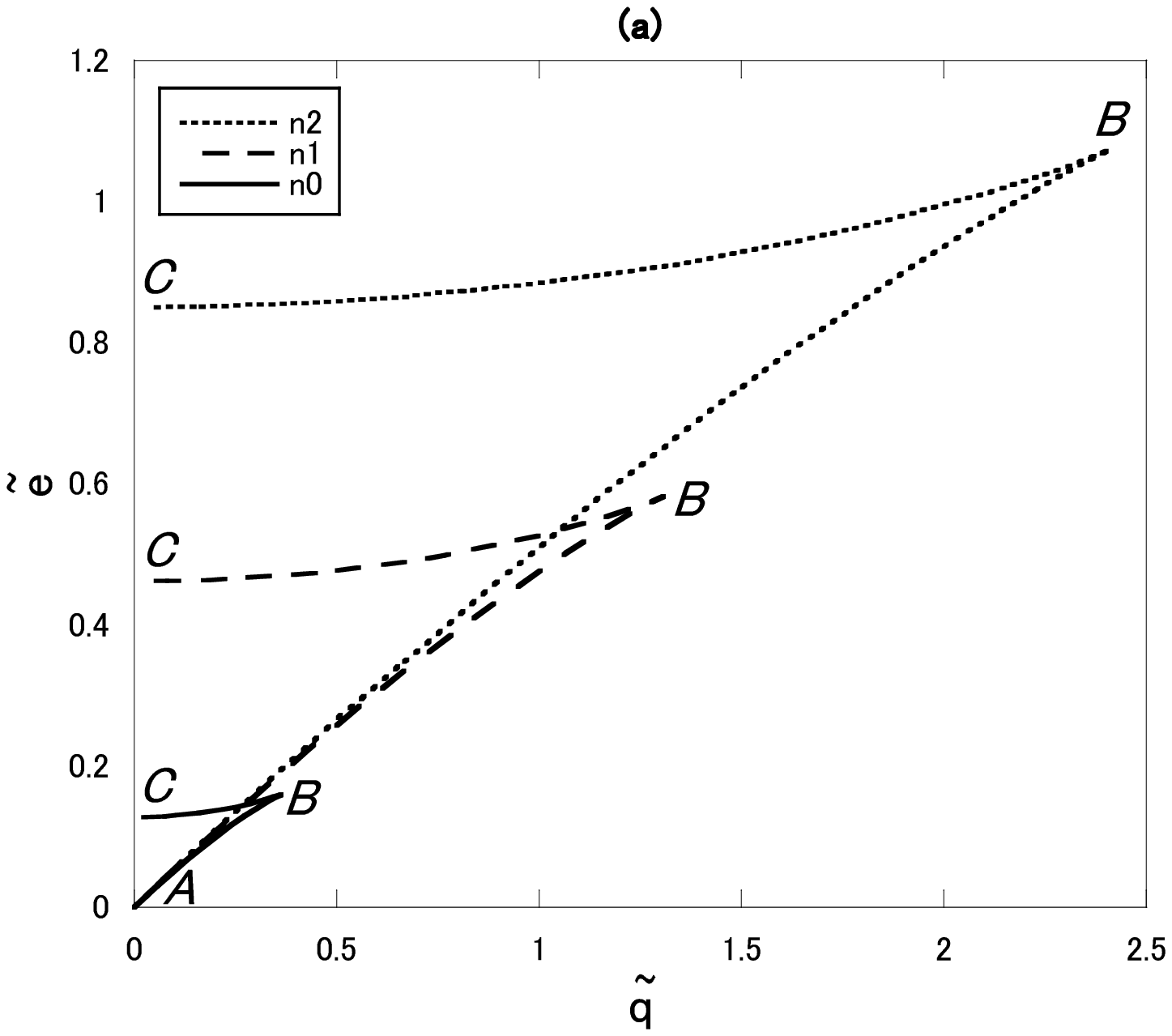,width=3.2in}
\psfig{file=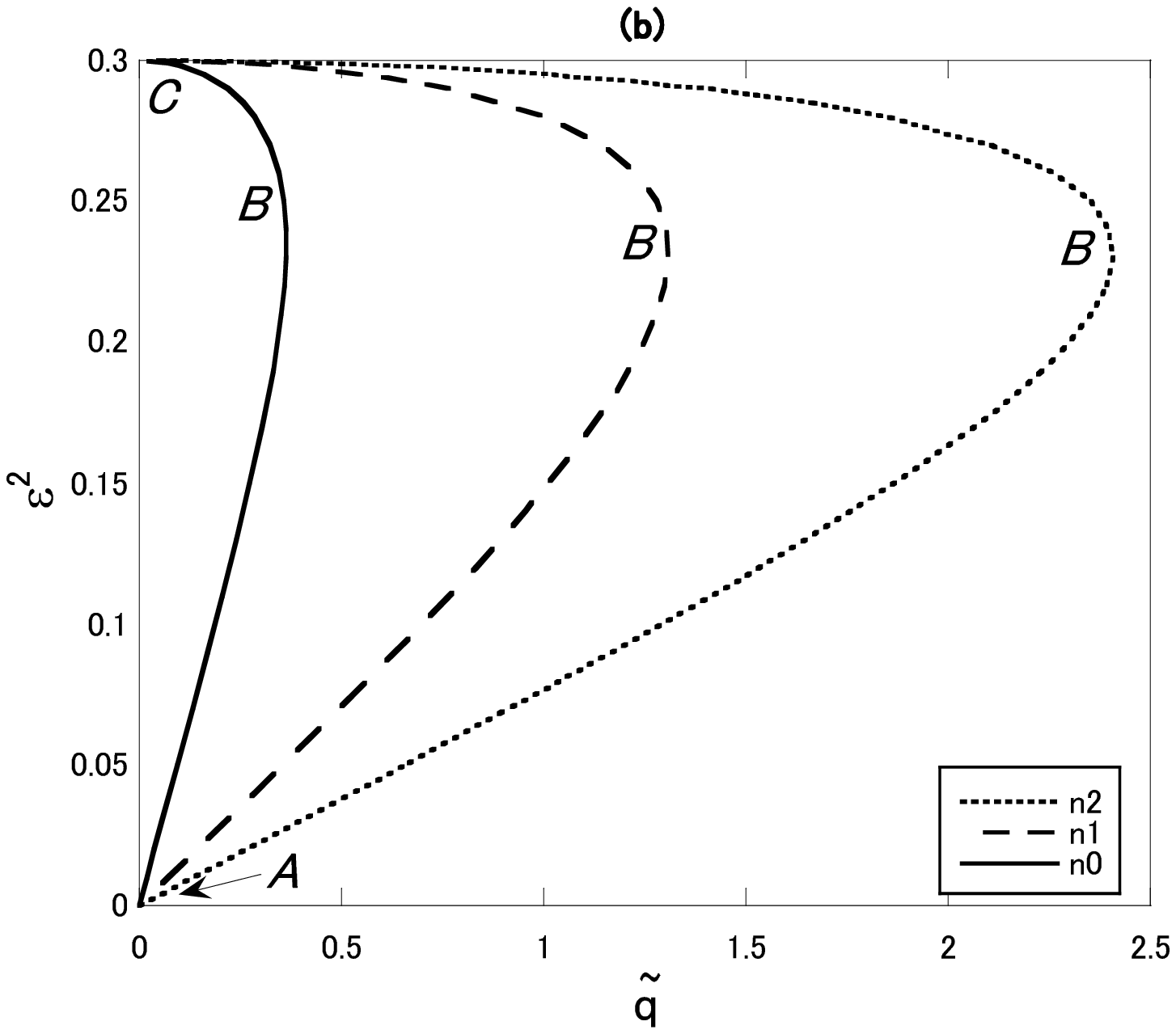,width=3.2in}
\caption{\label{qe-V3m03}
(a) $\tq$-$\te$ and (b) $\tq$-$\epsilon^2$ relations for Type I Q-tubes in the $V_{3}$ model:
$\tilde{m}^{2}=0.3$.}
\end{figure}

\section{Solutions in various potentials}

Here we investigate equilibrium solutions of Q-balls and Q-tubes for four types of potentials.

\subsection{$V_{3}$ model}

First, we summarize the previous results in the $V_{3}$ model (\ref{V3})~\cite{SIN}.
We rescale the quantities as
\bea
&&\tilde{\phi}\equiv{\lambda\over\mu}\phi,~
\tm\equiv{\sqrt{\lambda}\over\mu}m,~
\tilde{\omega}\equiv{\sqrt{\lambda}\over\mu}\omega , \nn
&&\tilde{r} \equiv \frac{\mu}{\sqrt{\lambda}}r,~\tilde{E} \equiv \frac{\lambda^{3/2}}{\mu}E,~
\tilde{Q} \equiv \lambda Q, \nn
&&\tilde{R} \equiv \frac{\mu}{\sqrt{\lambda}}R,~\te \equiv \frac{\lambda^2}{\mu^2}e,~
\tq \equiv \frac{\lambda^{3/2}}{\mu}q,
\label{rescale}
\eea
and define a parameter,
\beq\label{epsilonV3}
\epsilon^2\equiv\tm^2-\tilde{\omega}^2.
\eeq

Then, the existing condition (\ref{Qexist}) for the two types becomes
\bea
0<\epsilon^2<\frac12&{\rm for}&\tm^2>\frac12~({\rm Type~I})\nn
0<\epsilon^2<\tm^2&{\rm for}&\tm^2<\frac12~({\rm Type~II}).
\label{existV3}\eea
The limits $\epsilon^2\ra1/2$ and $\epsilon^2\ra0$ correspond to the thin-wall limit and the thick-wall limit, respectively.
As we discussed in the last section, however, in Type II solutions there is no thin-wall limit and the upper limit of $\epsilon^2$ is $\tm^2$ instead of $1/2$.

Figure \ref{field-V3} shows examples of the field configurations of Q-tubes.
We fix $\tm^2 =0.6$ (Type I), and choose $\epsilon^2=0.01$ (thick-wall) in (a) and $\epsilon^2=0.48$ (thin-wall) in (b). 
In each diagram we show the three solutions $n=0$, $1$ and $2$, which indicates that the maximum amplitude of the scalar field $\tilde{\phi}_{\rm max}$ for $n=0$ is largest among them.
We can understand it by analogy with the Newtonian mechanics in Fig.~\ref{Newton}. 
For $n\geq 1$, the particle must make a round trip while it goes an one-way for $n=0$. 
Nevertheless, $\tilde{\phi}_{\rm max}$ in all cases are qualitatively unchanged which 
means that the conservation law of energy approximately holds in words of the 
Newtonian mechanics. Of course, the behavior of a Q-ball is similar to that of a Q-tube for $n=0$.
These properties are independent of potentials, which is important in understanding Q-balls and Q-tubes in an unified way as we shall see in Sec. IV. 

We show the charge-energy-$\epsilon$ relations for Type I ($\tm^2=0.6$):
Q-balls in Fig.\ \ref{QEQball-V3m06} and Q-tubes in Fig.\ \ref{qe-V3m06}.
As for Q-tubes, we show results for $n=0$, $1$ and $2$. 
Similarity between Q-balls and Q-tubes is quite remarkable. 
In the thin-wall limit ($\epsilon^2\ra1/2$), we confirm that $\tilde{Q}$, $\tilde{E}$, $\tq$ and $\te$ diverge. In the thick-wall limit ($\epsilon^2\ra0$), on the other hand, these quantities approach zero.

We also show the same relations for Type II ($\tm^2 =0.3$):
Q-balls in Fig.~\ref{QEQball-V3m03} and Q-tubes in Fig.\ \ref{qe-V3m03}.
The crucial difference from Type I is that 
$\tilde{Q}$ and $\tq$ approach zero in the upper limit $\epsilon^2 \to \tm^2$ while 
$\tilde{E}$ and $\te$ have nonzero finite values corresponding to the points $C$. 
As a result, $\tilde{Q}$, $\tilde{E}$, $\tq$ and $\te$ have maximum values 
for intermediate value of $\epsilon^2$ corresponding to the points $B$ where cusp 
structures appear in Figs.~\ref{QEQball-V3m03} and \ref{qe-V3m03} (a). 
The stability of Q-balls and Q-tubes can be understood using catastrophe theory~\cite{PS78}. 
Solutions from the point $A$ to $B$ is stable while $B$ to $C$ unstable. 

The extreme values of the energy and the charge of Q-balls and Q-tubes in the $V_3$ model are summarized in Table II.

\begin{table}
\begin{tabular}{|c|c|c|}\hline
& $\epsilon^2\ra$min$[1/2,~\tm^2]$ & $\epsilon^2\ra0$ (thick)\\\hline
Type I: $\tm^2>1/2$ & $\tE,\tQ,\te,\tq\ra\infty$ & $\tE,\tQ,\te,\tq\ra0$ \\\hline
Type II: $\tm^2<1/2$ & $\tE,\te\ra$nonzero finite & $\tE,\tQ,\te,\tq\ra0$ \\
            & $\tQ,\tq\ra0$ & \\\hline
\end{tabular}
\caption{Extreme values of the energy and the charge of Q-balls and Q-tubes in the $V_3$ model.}
\end{table}


\begin{figure}[b]
\psfig{file=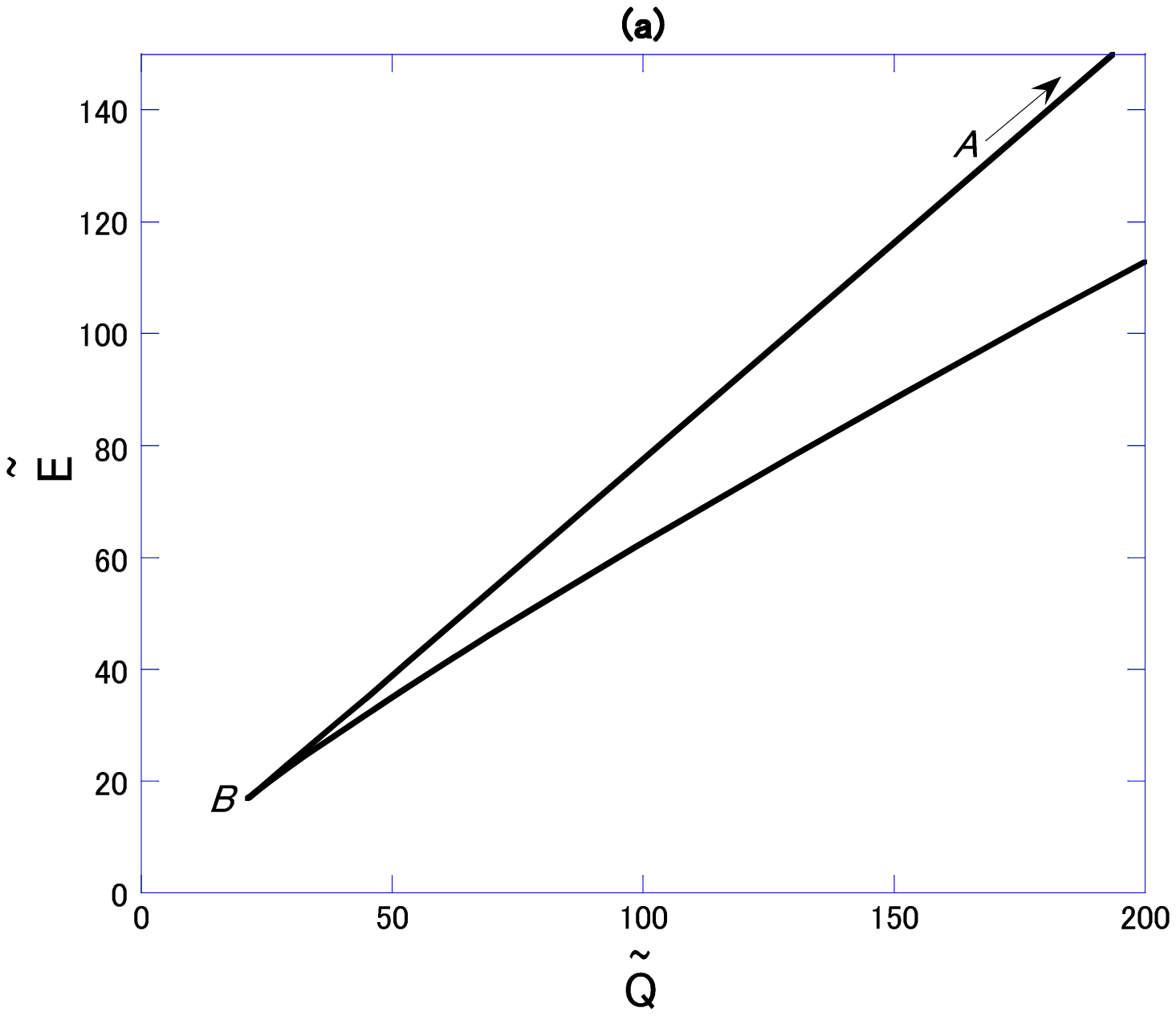,width=3.2in}
\psfig{file=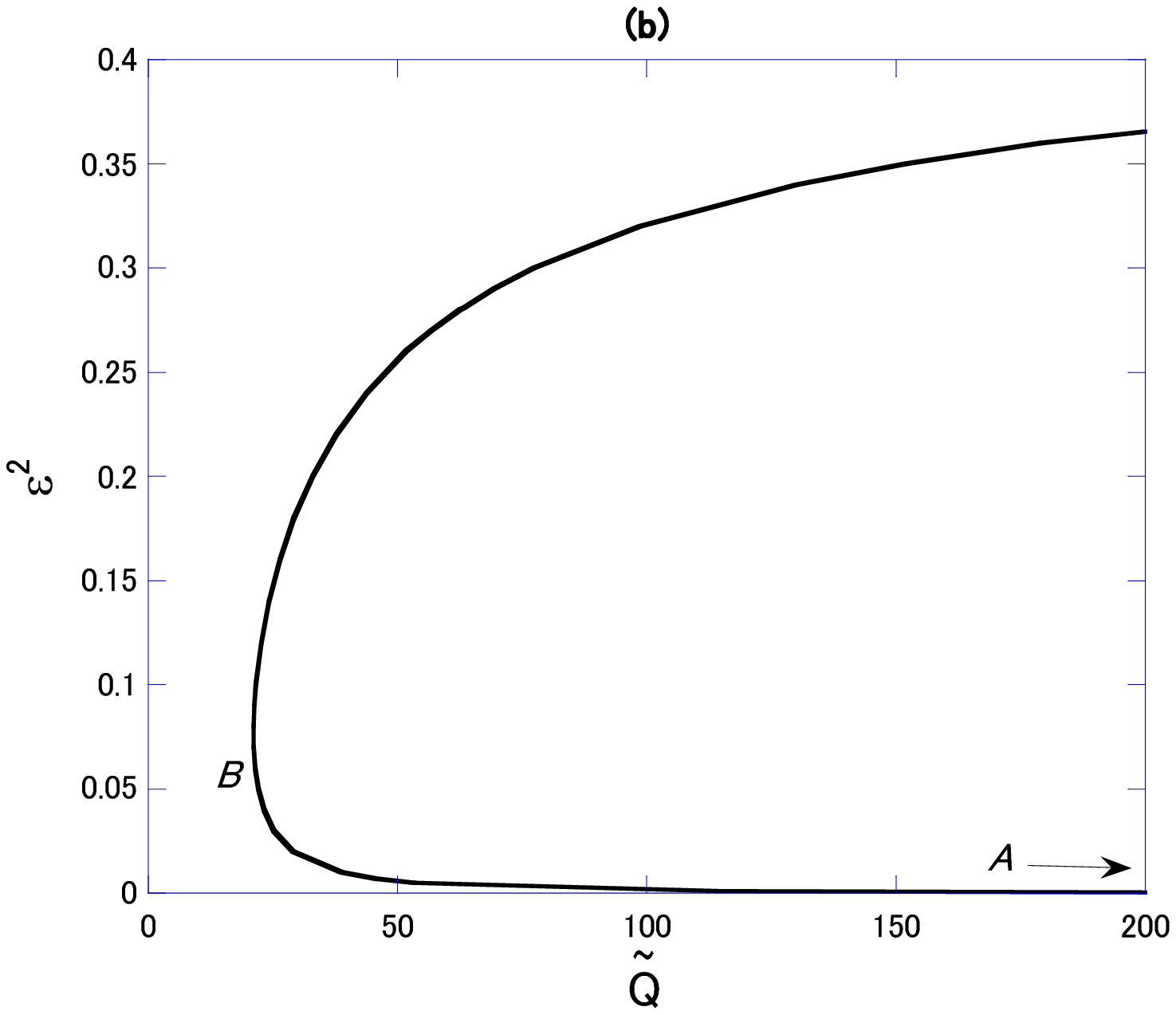,width=3.2in}
\caption{\label{QEQball-V4m06}
(a) $\tilde{Q}$-$\tilde{E}$ and (b) $\tilde{Q}$-$\epsilon^2$ relations for Type I Q-balls in the $V_{4}$ model: $\tilde{m}^{2}=0.6$.}
\end{figure}
\begin{figure}[b]
\psfig{file=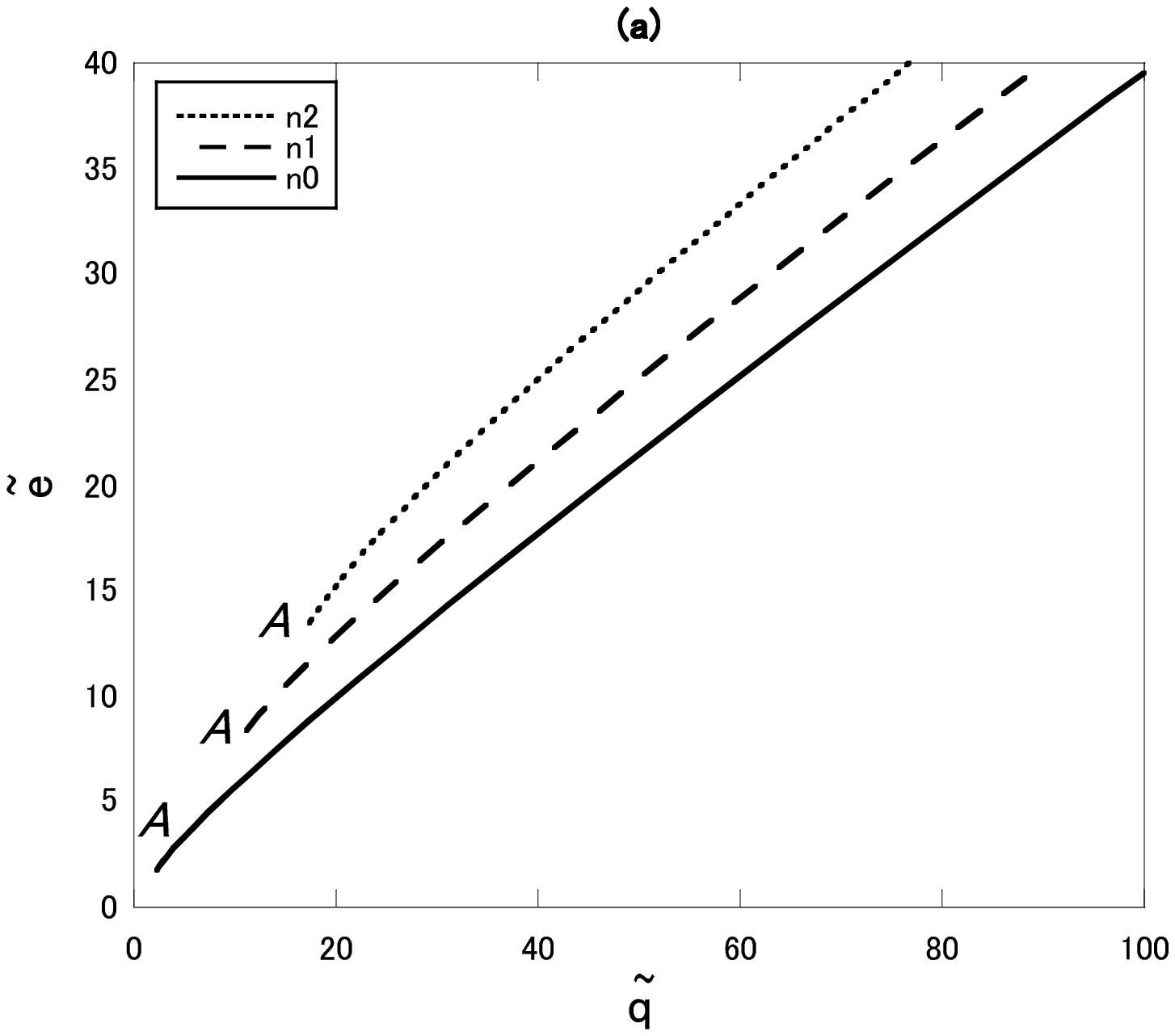,width=3.2in}
\psfig{file=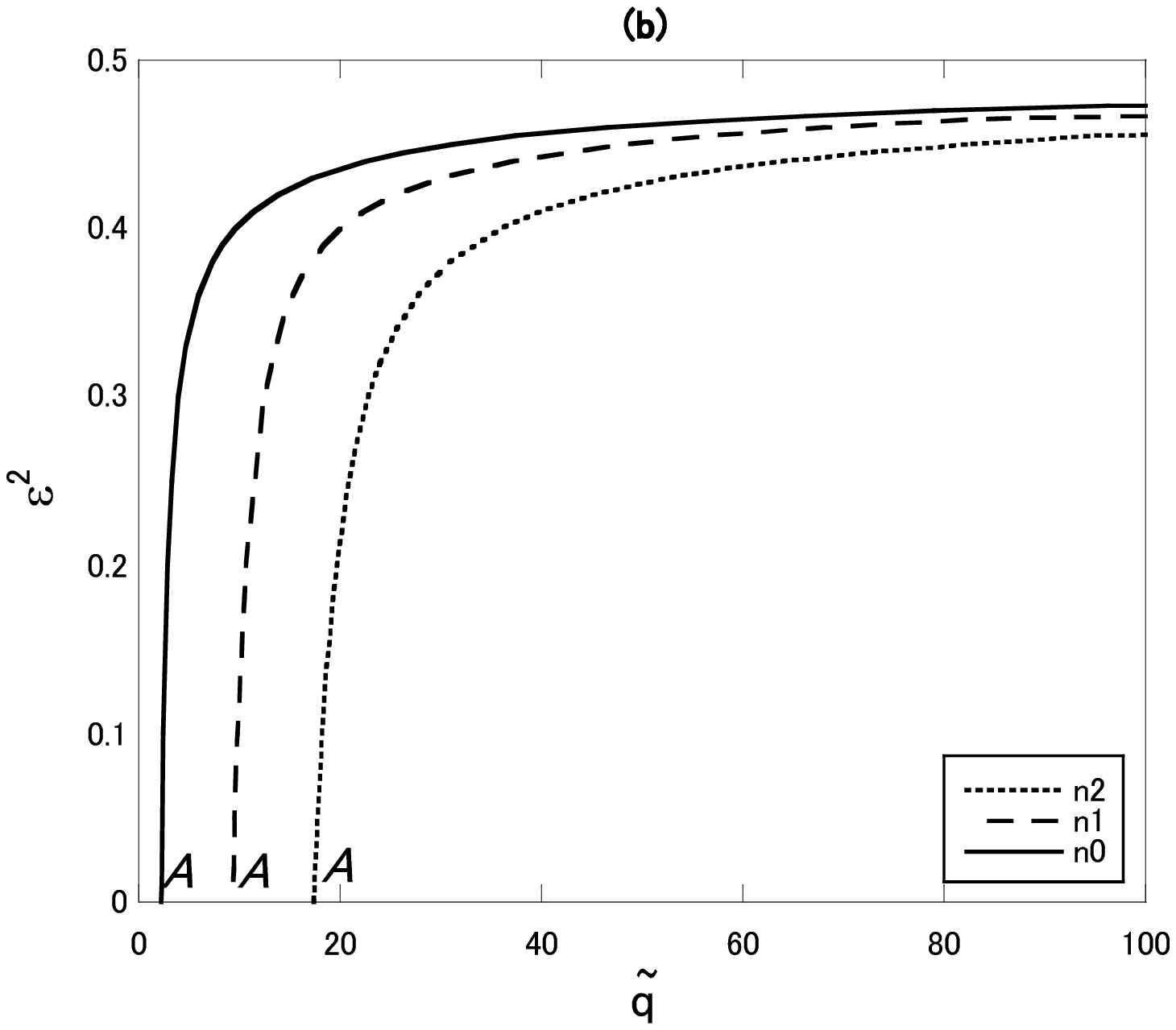,width=3.2in}
\caption{\label{qe-V4m06}
(a) $\tq$-$\te$ and (b) $\tq$-$\epsilon^2$ relations for Type I Q-tubes in the $V_{4}$ model:
$\tilde{m}^{2}=0.6$.}
\end{figure}
\begin{figure}[htbp]
\psfig{file=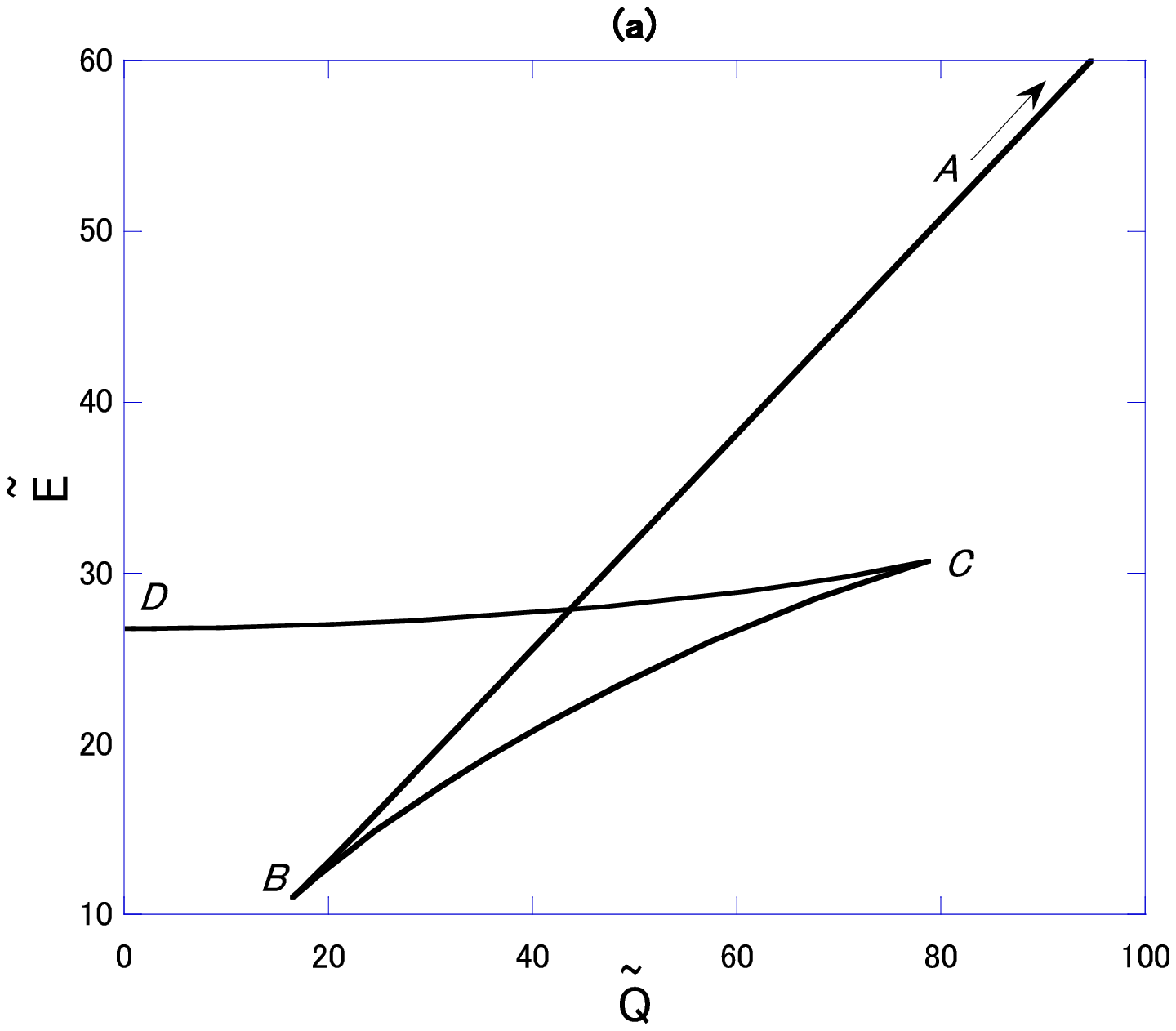,width=3.2in}
\psfig{file=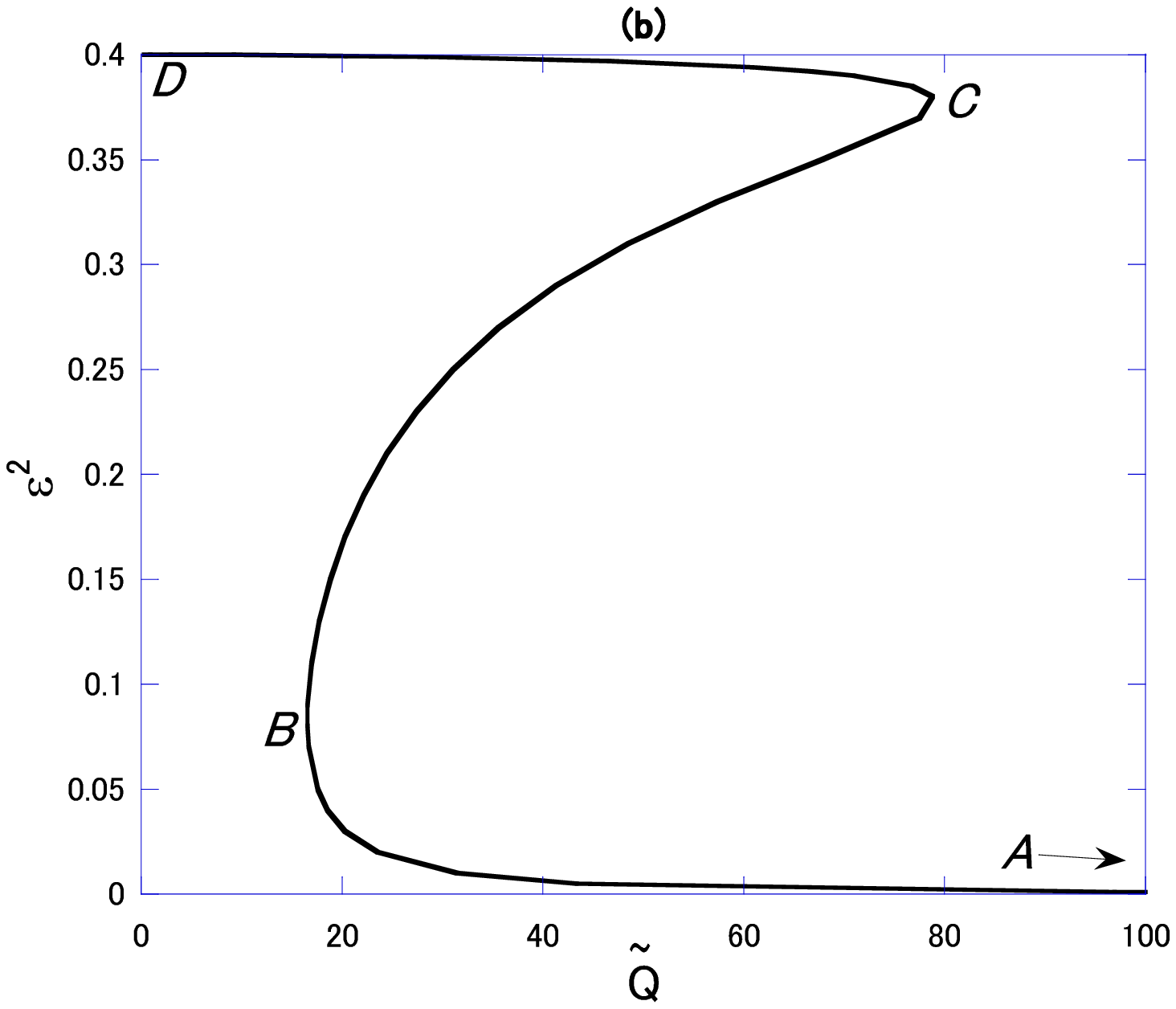,width=3.2in}
\caption{\label{QEQball-V4m04}
(a) $\tilde{Q}$-$\tilde{E}$ and (b) $\tilde{Q}$-$\epsilon^2$ relations for Type II Q-balls in the $V_{4}$ model: $\tilde{m}^{2}=0.4$.}
\end{figure}
\begin{figure}[htbp]
\psfig{file=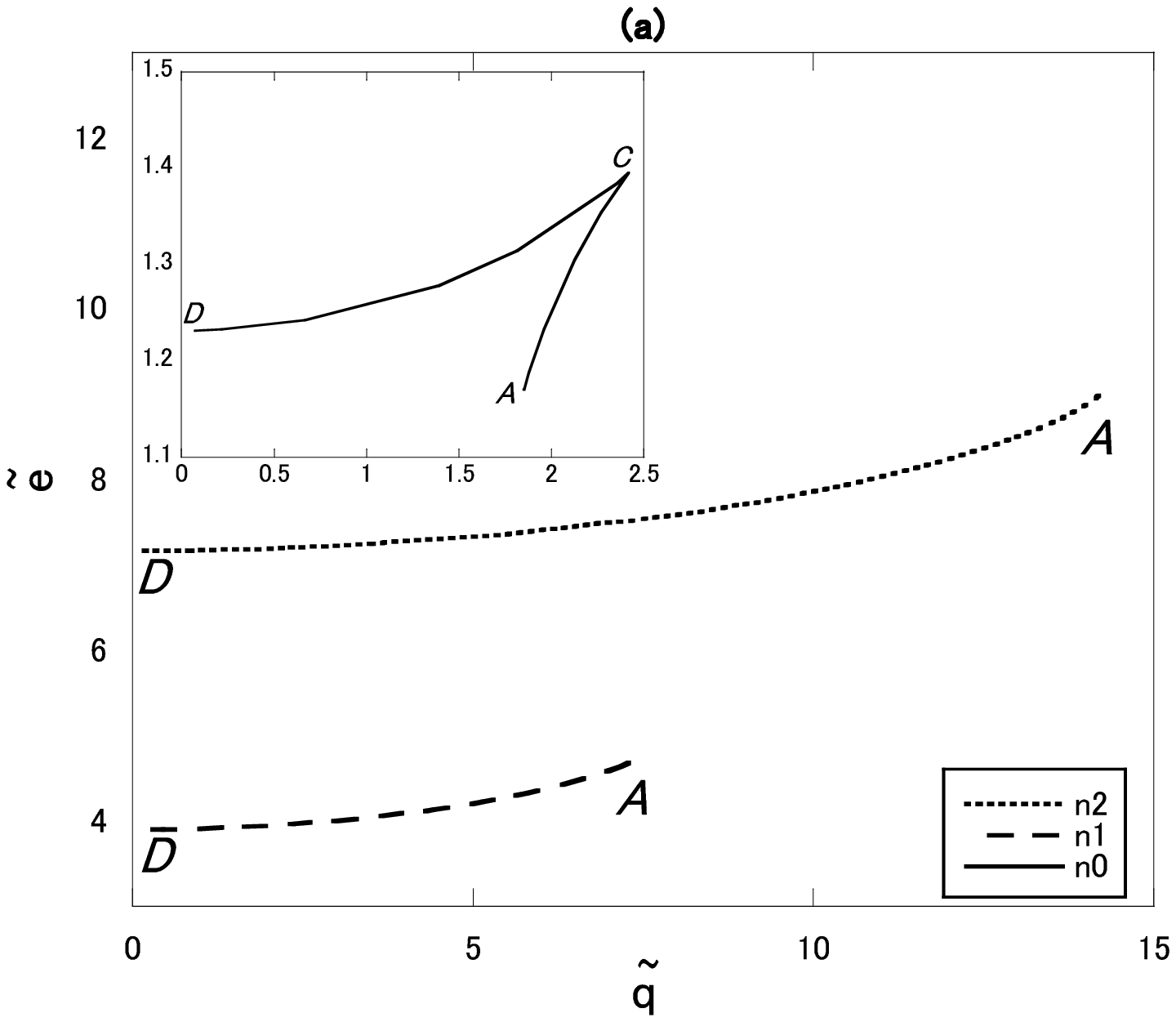,width=3.2in}
\psfig{file=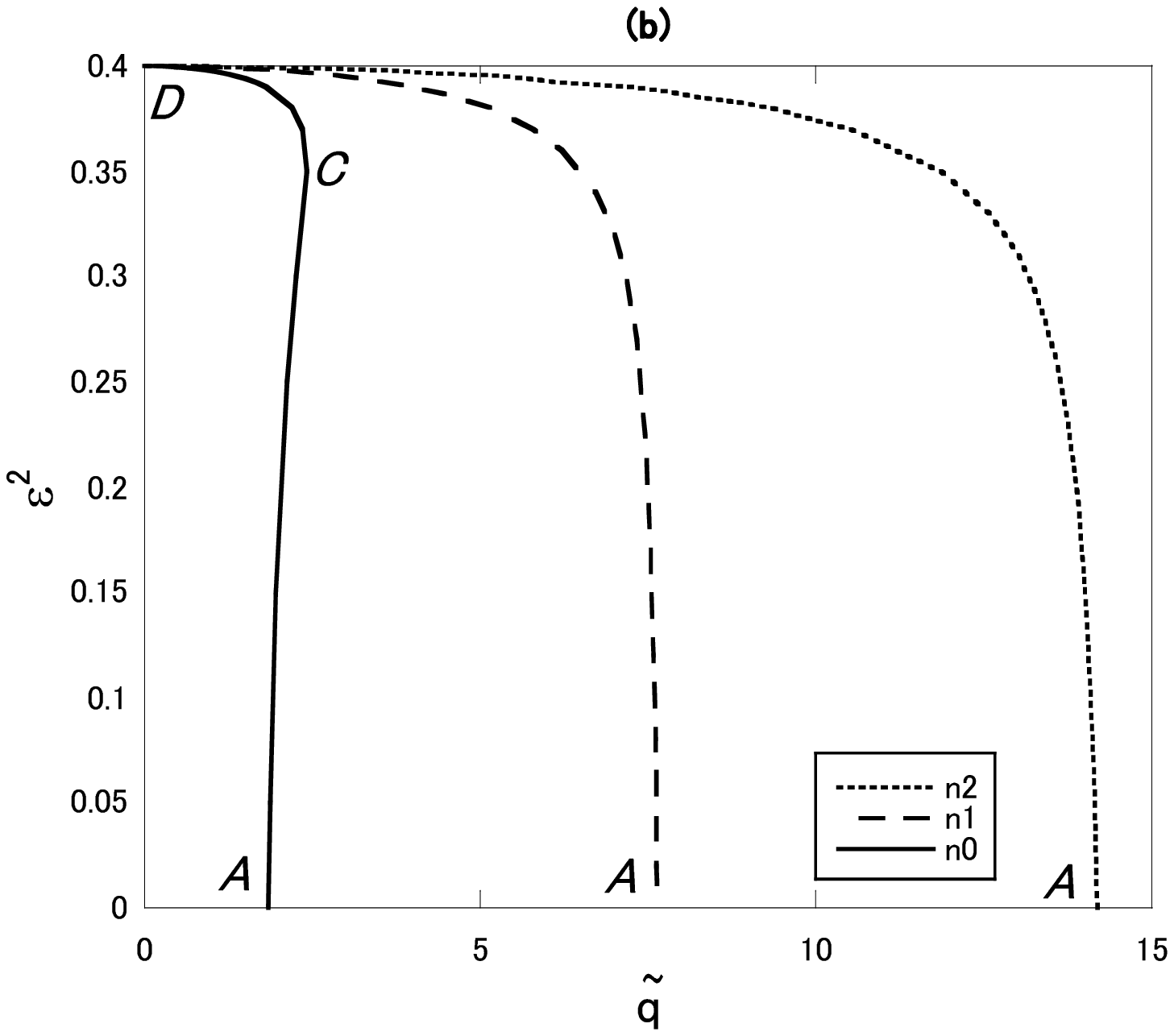,width=3.2in}
\caption{\label{qe-V4m04}
(a) $\tq$-$\te$ and (b) $\tq$-$\epsilon^2$ relations for Type II Q-tubes in the $V_{4}$ model:
$\tilde{m}^{2}=0.4$.}
\end{figure}

\subsection{the $V_{4}$ model}

Second, we consider another simple potential,
\beq\label{V4}
V_4(\phi):={m^2\over2}\phi^2-\lambda\phi^4+\frac{\phi^6}{M^2} 
~~~{\rm with} ~~~ m^2,~\lambda,~M^2>0,
\eeq
which we call the $V_4$ model.
We rescale the quantities as
\bea
&&\tp \equiv{\phi\over\sqrt{\lambda}M},~~\tm \equiv{m\over\lambda M},~~
\tilde{\omega} \equiv{\omega\over\lambda M}, \nonumber  \\
&&\tilde{r} \equiv \lambda Mr,~~\tilde{E} \equiv {E\over M},~~ \tilde{Q} \equiv \lambda Q,
\nonumber  \\
&&\tilde{R} \equiv \lambda MR,~~\te \equiv \frac{e}{\lambda M^2}~~\tq \equiv\frac{q}{M},
\label{rescaleV4}
\eea
and again define a parameter $\epsilon$ by (\ref{epsilonV3}).

Then the existing condition is identical to (\ref{existV3}) in the $V_3$ case.
We show the charge-energy-$\epsilon$ relations in Figs.\ \ref{QEQball-V4m06}-\ref{qe-V4m04}:
Type I Q-balls in Fig.\ \ref{QEQball-V4m06}, Type I Q-tubes in Fig.\ \ref{qe-V4m06},
Type II Q-balls in Fig.\ \ref{QEQball-V4m04}, and Type II Q-tubes in Fig.\ \ref{qe-V4m04}.
Contrary to the case of the $V_{3}$ model, qualitative difference between Q-tubes and Q-balls appears.
The extreme values of the energy and the charge of Q-balls and Q-tubes in the $V_4$ model are summarized in Table III.

\begin{table}
\begin{tabular}{|c|c|c|}\hline
& $\epsilon^2\ra$min$[1/2,~\tm^2]$ & $\epsilon^2\ra0$ (thick)\\\hline
Type I: $\tm^2>1/2$ & $\tE,\tQ,\te,\tq\ra\infty$ & $\tE,\tQ\ra\infty$ \\
                                 &                                     & $\te,\tq\ra$nonzero finite\\\hline
Type II: $\tm^2<1/2$ & $\tE,\te\ra$nonzero finite & $\tE,\tQ\ra\infty$ \\
                                 & $\tQ,\tq\ra0$ & $\te,\tq\ra$nonzero finite \\\hline
\end{tabular}
\caption{Extreme values of the energy and the charge of Q-balls and Q-tubes in the $V_4$ model.}
\end{table}

The structures of the solution series of Type II Q-balls and Q-tubes are not simple.
In the case of Q-balls, there are two cusps in the $Q$-$E$ diagram, $B$ and $C$.
Only the solutions between these two points represent stable solutions.
In the case of Q-tubes, a cusp appears for $n=0$, while no cusp appears for $n\ge1$.


\begin{figure}[htbp]
\psfig{file=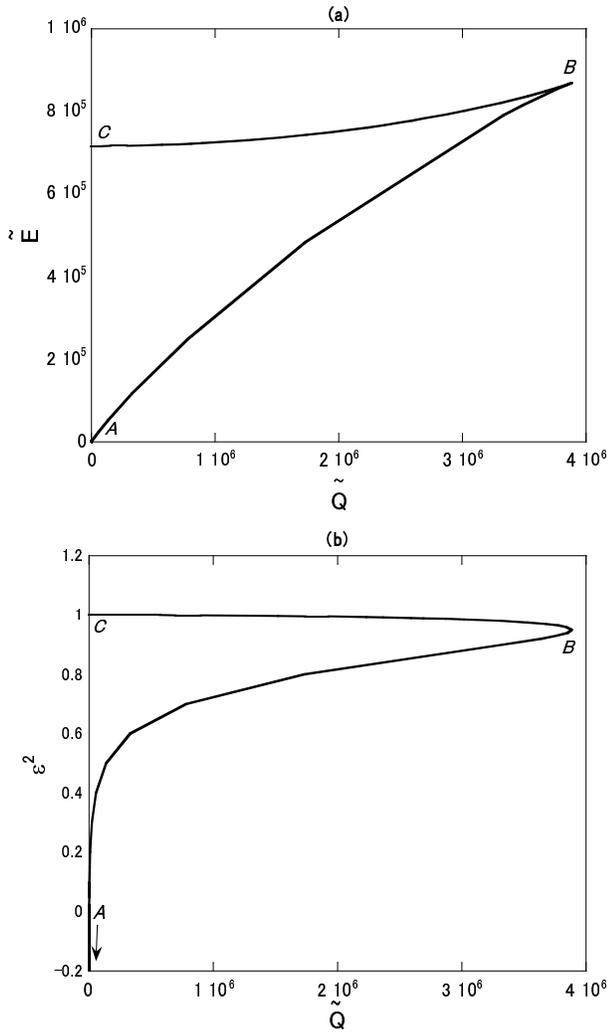,width=3.2in}
\psfig{file=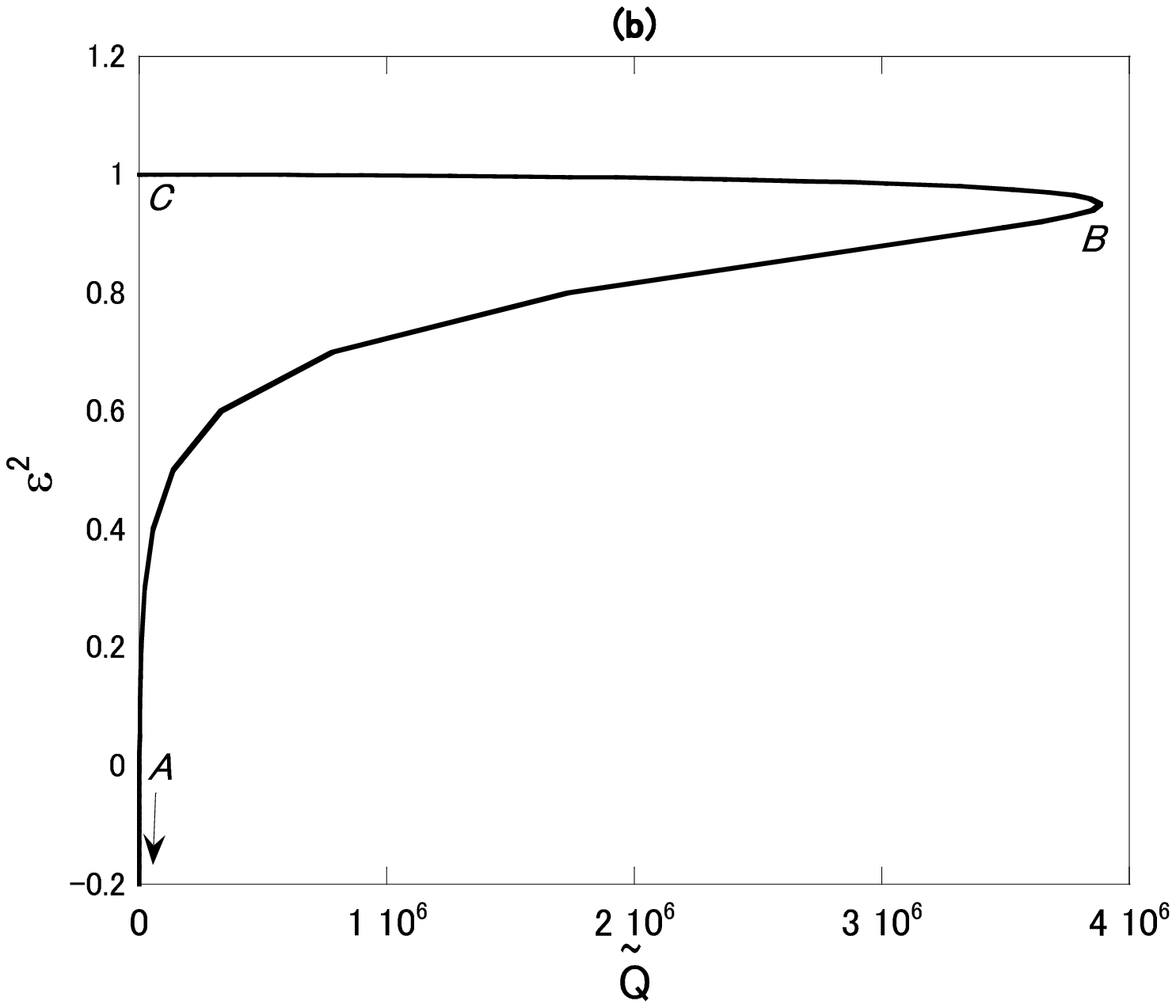,width=3.2in}
\caption{\label{QEQball-gravityK-01}
(a) $\tilde{Q}$-$\tilde{E}$ and (b) $\tilde{Q}$-$\epsilon^2$ relations for $V_{\rm grav.}$: $K=-0.1$.}
\end{figure}
\begin{figure}[htbp]
\psfig{file=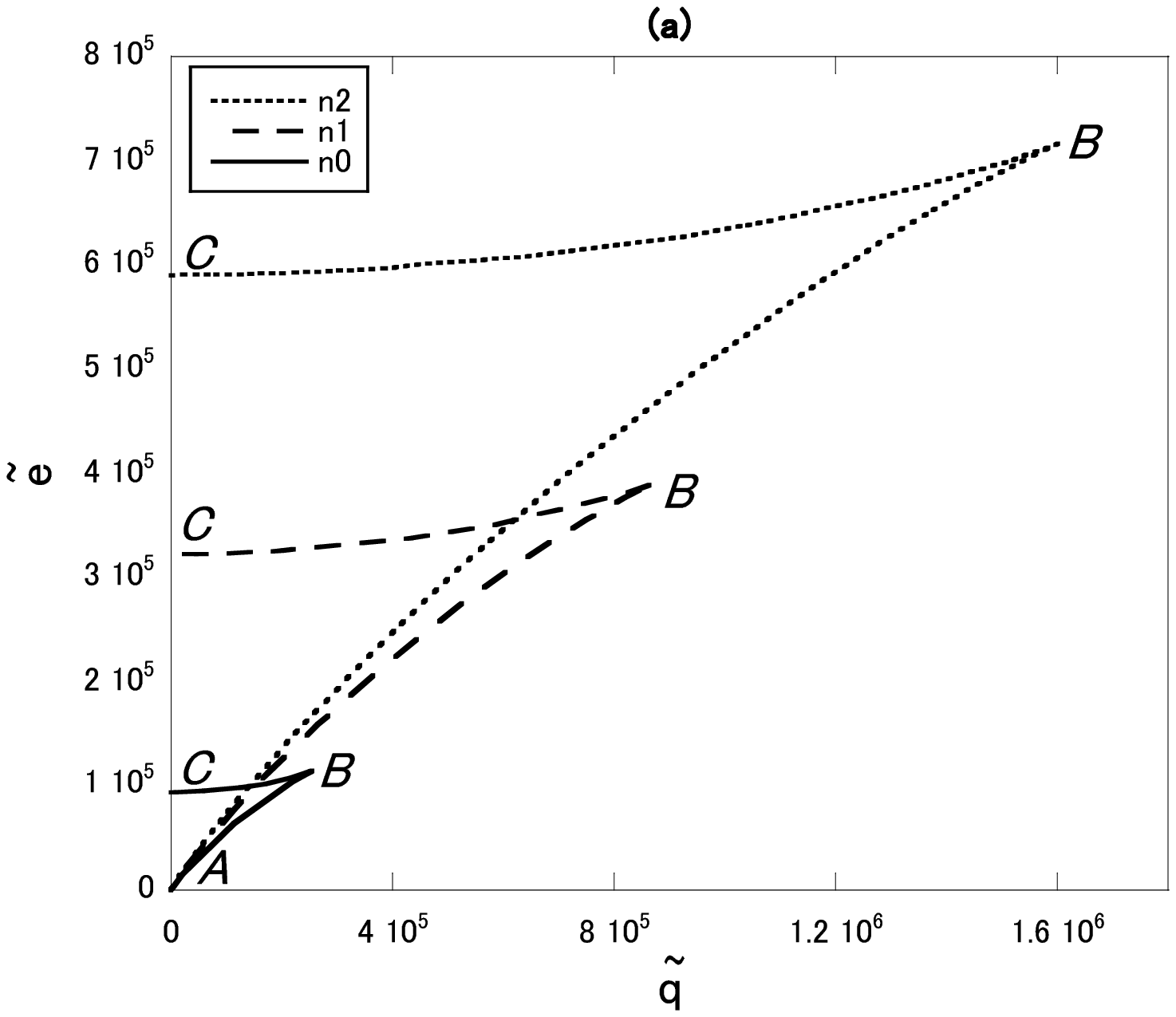,width=3.2in}
\psfig{file=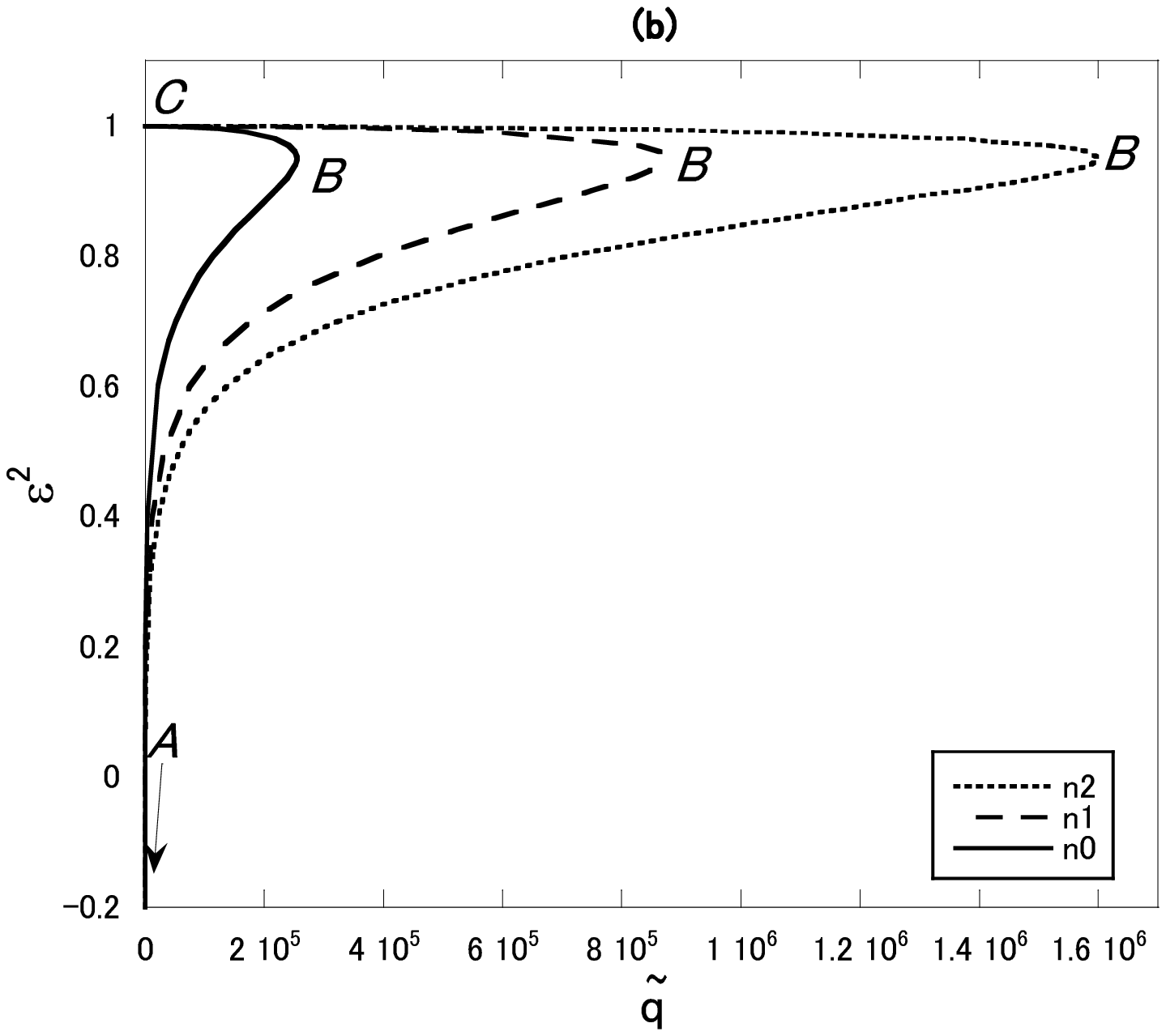,width=3.2in}
\caption{\label{qe-gravity}
(a) $\tq$-$\te$ and (b) $\tq$-$\epsilon^2$ relations for $V_{\rm grav.}$: $K=-0.1$.}
\end{figure}

\subsection{AD gravity-mediation type}

From the theoretical point of view, it is important to investigate Q-tubes as well as Q-balls 
in the AD mechanism. 
There are two types of potentials: gravity-mediation type and gauge-mediation type. 
Here we consider the former type,
\bea\label{gravity}
&&V_{\rm grav.}(\phi):=\frac{m_{\rm grav.}^2}{2}\phi^2\left[
1+K\ln \left(\frac{\phi}{M}\right)^2
\right]~~  \nonumber  \\
&&{\rm with} ~~ m_{\rm grav.}^2,~M>0.
\eea
We rescale the quantities as 
\bea
&&\tp\equiv\frac{\phi}{M},~~\tilde{\omega}\equiv\frac{\omega}{m_{\rm grav.}},~~ \nonumber  \\
&&\tilde{r}\equiv m_{\rm grav.}r, ~~\tilde{E}\equiv \frac{m_{\rm grav.}E}{M^2},
~~ \tilde{Q}\equiv \frac{m_{\rm grav.}^2 Q}{M^2}, 
\nonumber \\
&&\tilde{R}\equiv m_{\rm grav.}R, 
~~\te\equiv \frac{e}{M^2},~~ \tq\equiv \frac{m_{\rm grav.} q}{M^2},
\label{rescale-gravity}
\eea
and define a parameter $\epsilon$ as 
\bea
\epsilon^2 =1-\tilde{\omega}^2.
\label{epsilon-gravity}
\eea

The existing condition (\ref{Qexist}) becomes
\beq
K<0,~~~ \epsilon^2<1.
\label{gravity-condition}
\eeq
Thus, $\epsilon^2$ is not bounded below, which is in contrast to the $V_{3}$ and $V_{4}$ models.
Only Type II solutions exist in this model unless we introduce additional terms in the potential.
We show the charge-energy-$\epsilon$ relations:
Q-balls in Fig.\ \ref{QEQball-gravityK-01} and Q-tubes in Fig.\ \ref{qe-gravity}.
The extreme values of the energy and the charge of Q-balls and Q-tubes in the gravity-mediation type are summarized in Table IV.
There is no qualitative difference in the charge-energy relation between Q-balls and Q-tubes.
These properties are common to Type II solutions in the $V_{3}$ model.

\begin{table}
\begin{tabular}{|c|c|c|}\hline
& $\epsilon^2\ra1$ & $\epsilon^2\ra-\infty$ (thick)\\\hline
Type II & $\tE,\te\ra$nonzero finite & $\tE,\tQ,\te,\tq\ra0$ \\
            & $\tQ,\tq\ra0$ &  \\\hline
\end{tabular}
\caption{Extreme values of the energy and the charge of Q-balls and Q-tubes in the AD gravity-mediation type.}
\end{table}


\begin{figure}[htbp]
\psfig{file=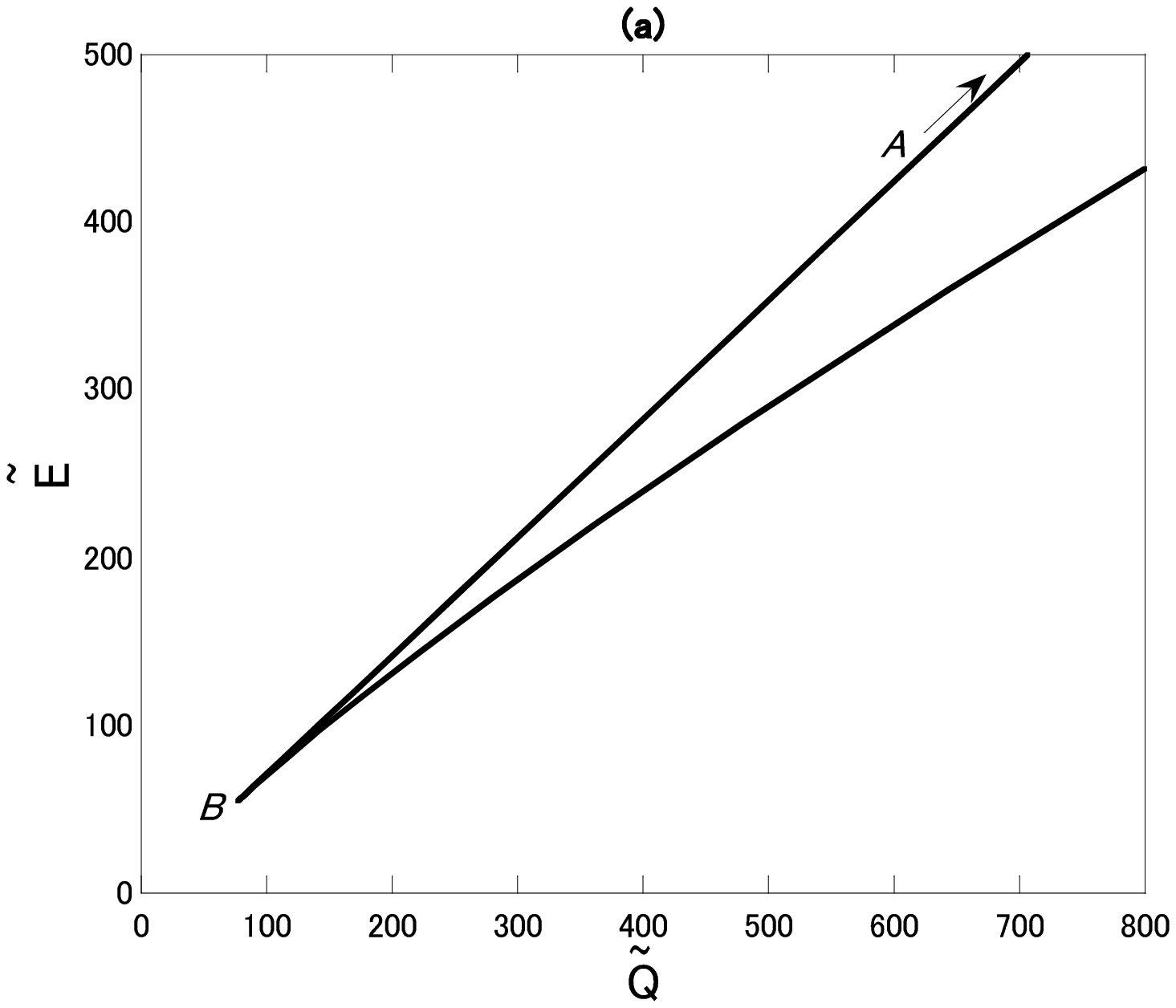,width=3.2in}
\psfig{file=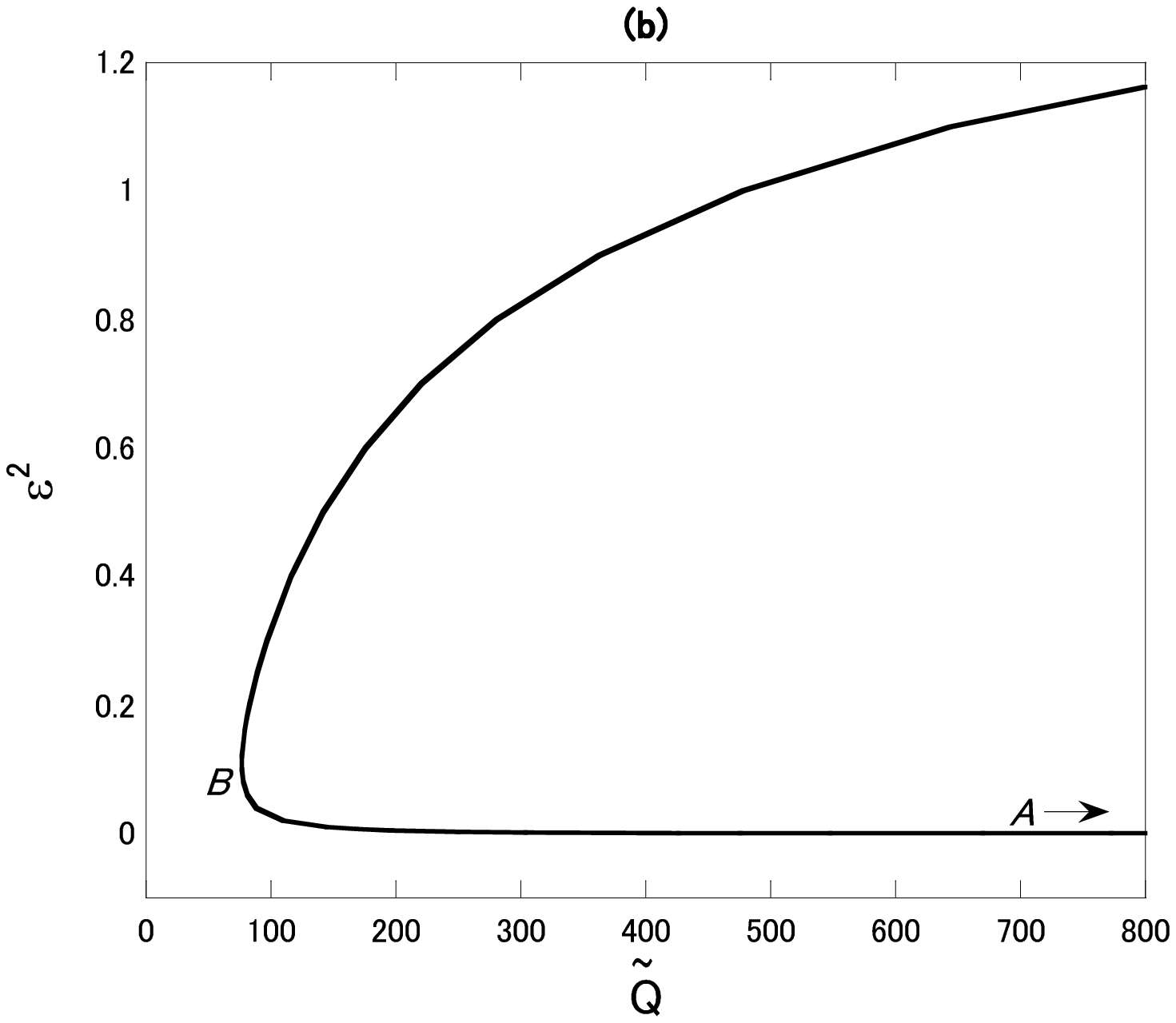,width=3.2in}
\caption{\label{QEQball-gauge}
(a) $\tilde{Q}$-$\tilde{E}$ and (b) $\tilde{Q}$-$\epsilon^2$ relations for $V_{\rm gauge}$.}
\end{figure}
\begin{figure}[htbp]
\psfig{file=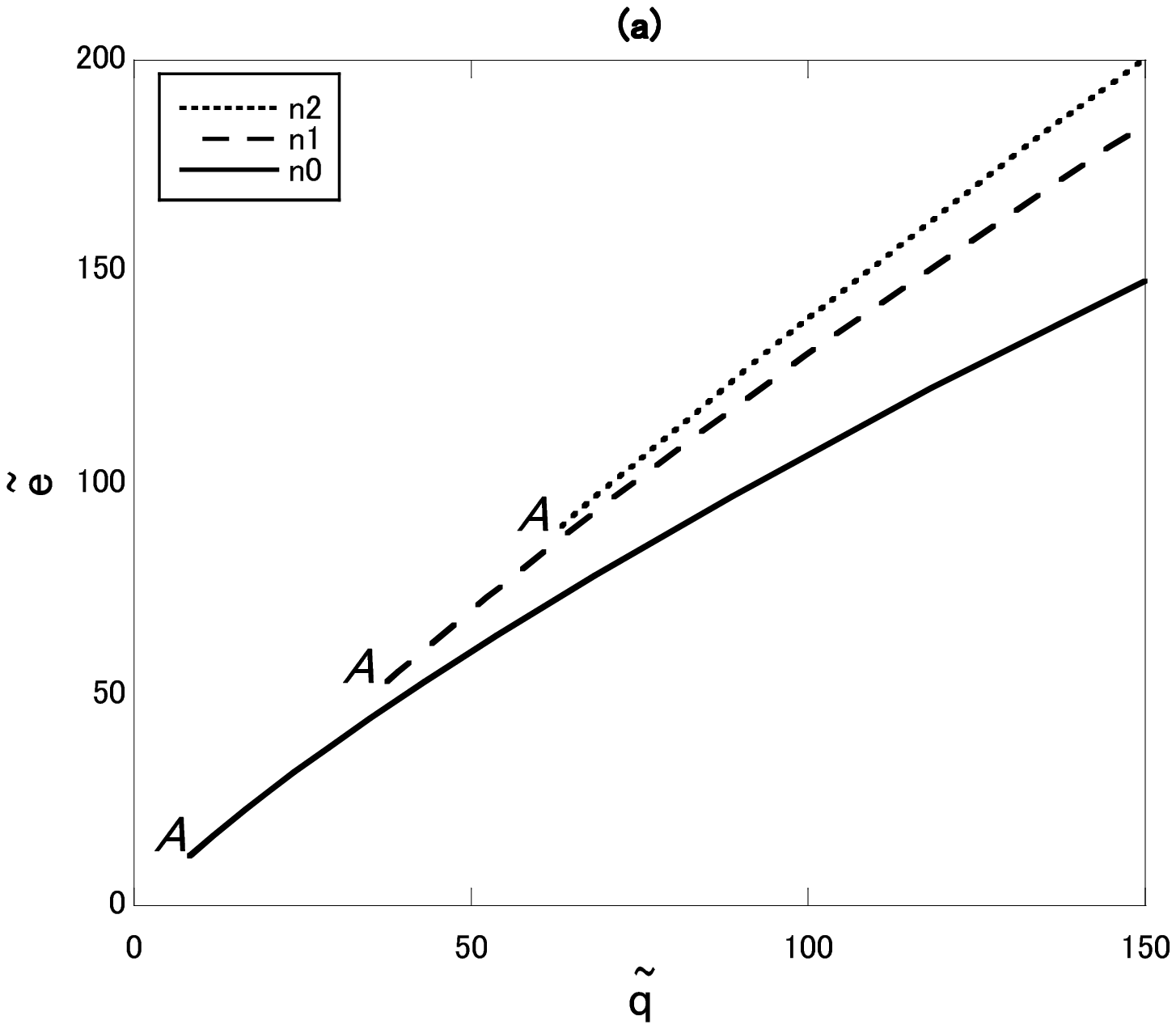,width=3.2in}
\psfig{file=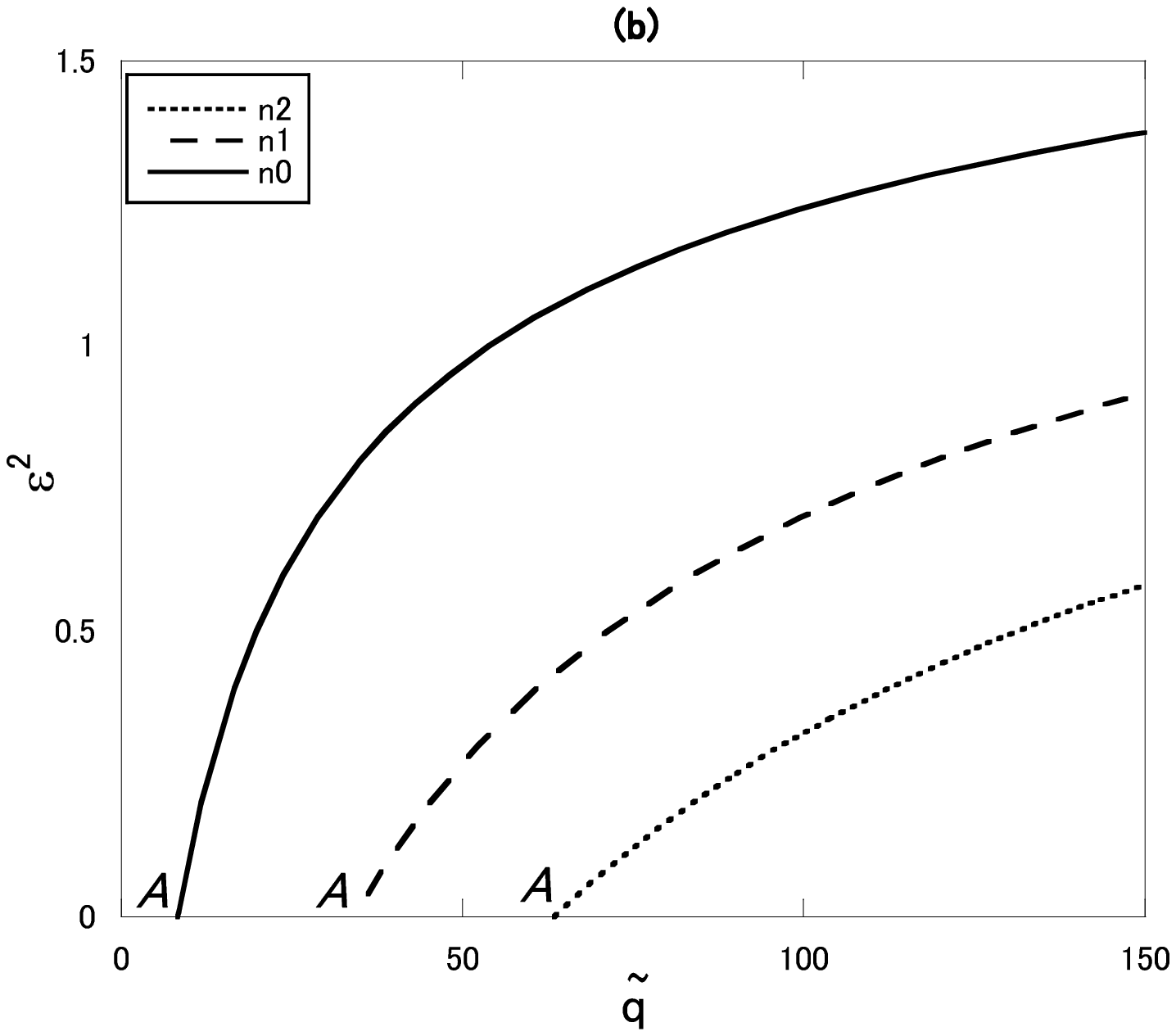,width=3.2in}
\caption{\label{qe-gauge}
(a) $\tq$-$\te$  and (b) $\tq$-$\epsilon^2$ relations for $V_{\rm gauge}$.}
\end{figure}

\subsection{AD gauge-mediation type}

Finally, we consider the gauge-mediation type in the AD mechanicsm,
\beq\label{gauge}
V_{\rm gauge}(\phi):=m_{\rm gauge}^4 \ln\left(1+\frac{\phi^2}{m_{\rm gauge}^2}\right)~~~
{\rm with} ~~~ m_{\rm gauge}^2>0\ .
\eeq
We rescale the quantities as 
\bea
&&\tp\equiv\frac{\phi}{m_{\rm gauge}},~~\tilde{\omega}\equiv\frac{\omega}{m_{\rm gauge}}, 
\nonumber  \\
&&\tilde{r}\equiv m_{\rm gauge}r,~~\tilde{E}\equiv \frac{E}{m_{\rm gauge}},~~\tilde{Q}\equiv Q, \nonumber  \\
&&\tilde{R}\equiv m_{\rm gauge}R,~~\te\equiv \frac{e}{m_{\rm gauge}^2},~~\tq\equiv \frac{q}{m_{\rm gauge}},
\label{rescale-gauge}
\eea
and define a parameter $\epsilon$ as 
\beq
\epsilon^2 =2-\tilde{\omega}^2.
\label{epsilon2}
\eeq

Then the existing condition (\ref{Qexist}) becomes
\beq
0<\epsilon^2<2 .
\label{gauge-condition}
\eeq
Only Type I solutions exist in this model.
We show the charge-energy-$\epsilon$ relation:
Q-balls in Fig.\ \ref{QEQball-gauge} and Q-tubes in Fig.\ \ref{qe-gauge}.
The extreme values of the energy and the charge of Q-balls and Q-tubes in the gravity-mediation type are summarized in Table V.
These properties are common to the Type I solutions in the $V_{4}$ model.

\begin{table}
\begin{tabular}{|c|c|c|}\hline
           & $\epsilon^2\ra2$ (thin) & $\epsilon^2\ra0$ (thick)\\\hline
Type I & $\tE,\tQ,\te,\tq\ra\infty$ & $\tE,\tQ\ra\infty$ \\
            &                                     &  $\te,\tq\ra$nonzero finite \\\hline
\end{tabular}
\caption{Extreme values of the energy and the charge of Q-balls and Q-tubes in the AD gauge-mediation type.}
\end{table}


\section{Unified picture of Q-balls and Q-tubes}

Our numerical results in the last section indicate that the charge-energy relation of equilibrium solutions depends a great deal on functional forms of the potential $V(\phi)$.
In this section we discuss what determines the extreme values of the energy and the charge by analytical methods.
As we explained in Sec. II, we can understand Q-balls and Q-tubes in words of a particle motion in Newtonian mechanics.
In Fig.~\ref{Newton}, if we ignore \lq non conserved force\rq\, the maximum of $\phi$, $\tilde{\phi}_{\rm max}$, is determined by the nontrivial solution of $V_{\omega}=0$. 
Using this $\tp_{\rm max}$, we can evaluate the order of magnitude of the energy and the charge, (\ref{EQball-definition}) and (\ref{eq-definition}), as 
\bea
\tilde{E}&\sim &\tilde{r}_{\rm max}^3
\left\{{1\over2}\tilde{\omega}^2\tilde{\phi}_{\rm max}^2+
\frac12\left({d\tilde{\phi}\over d\tilde{r}}\right)^2
+\tilde{V}\right\},\nn
\tilde{Q} &\sim &\tilde{\omega} \tilde{r}_{\rm max}^3 \tilde{\phi}_{\rm max}^2, \nn
\te&\sim &\tilde{R}_{\rm max}^2
\left\{{1\over2}\tilde{\omega}^2\tilde{\phi}_{\rm max}^2+
\frac12\left({d\tilde{\phi}\over d\tilde{R}}\right)^2
+\frac{n^2\tilde{\phi}_{\rm max}^2}{2\tilde{R}_{\rm max}^2}
+\tilde{V}\right\},\nn
\tq &\sim &\tilde{\omega} \tilde{R}_{\rm max}^2 \tilde{\phi}_{\rm max}^2,
\label{tq-evaluate}
\eea
where the subscript ``max" denote the values at which $\tp=\tp_{\rm max}$.
As for $\tilde{R}_{\rm max}$ for $n=0$ or $\tilde{r}_{\rm max}$, 
it is reasonable to take 
$\tilde{R}$ or $\tilde{r}$ where $\tilde{\phi}$ becomes about $0.5\tilde{\phi}_{\rm max}$. 

What we want to discuss is whether $\tE,~\tQ,~\te$ and $\tq$ approach zero, infinity or nonzero finite values as $\epsilon^2$ approaches the upper or lower limit.
The approximate expression (\ref{tq-evaluate}) is appropriate for this purpose.



First, we discuss the upper limit of $\epsilon^2$, or equivalently, the lower limit of $\omega^2$.
In Type I solutions, where min$[V]=V(0)=0$, in the limit of $\omega\ra$min$[2V/\phi^2]$, the minimum of $V_{\omega}$ approaches zero.
In this case, in the Newtonian-mechanics picture of Fig.\ 1, a particle rolls down from the top of the hill over infinite time, i.e., $R_{\rm max}$ diverges.
This limit corresponds to the thin-wall limit.
From the expression (\ref{tq-evaluate}), we see that $\tilde{Q}$, $\tilde{E}$, $\tq$ and $\te$ diverge.

On the other hand, in the Type II solutions, where min$[V]<0$, because $V_{\omega}<V$, there is no limit of min$V_{\omega}\ra0$.
Therefore, $\tilde{Q}$, $\tilde{E}$, $\tq$ and $\te$ must have their upper limits.

Next, we investigate the lower limit of $\epsilon^2$, or equivalently, the upper limit of $\omega^2$.
This limit corresponds to the thick-wall limit.
Except for the $V_{\rm grav.}$ model, $\epsilon$ satisfies
\beq
\epsilon^2=\frac{1}{m^2}{d^2V_{\omega}\over d\phi^2}(0),
\eeq
which means that $\epsilon$ is the mass scale of $V_{\omega}$ normalized by $m$.
Therefore, the wall thickness normalized by $m$ is of order of $1/\epsilon$.
Because the radius and the wall thickness are of the same order in the thick-wall limit, 
except for the $V_{\rm grav.}$ model, we obtain
\bea
\tilde{r}_{\rm max},~~\tilde{R}_{\rm max}\sim \frac{1}{\epsilon}.
\label{Rmax-evaluate}
\eea

In the following, from the approximate expression (\ref{tq-evaluate}) and (\ref{Rmax-evaluate}) we evaluate the limits of the charge and the energy as $\epsilon$ approaches the lower limit.

\vspace{5mm}

{\bf (A) $V_{3}$ case}

The solution of $V_{\omega}=0$ is 
\bea
\tilde{\phi}_{\rm max}=\frac{1- \sqrt{1-2\epsilon^2}}{2} .
\label{V3-solution}
\eea
In the lower limit $\epsilon^2 \to 0$, we have $\tilde{\phi}_{\rm max}\simeq{\epsilon^2}/{2}$. 
Therefore, from (\ref{tq-evaluate})-(\ref{Rmax-evaluate}), we find
\beq
\tQ,~\tE,~\tq,~\te\ra0,
\eeq
which agree with the numerical results in Table II.

\vspace{5mm}

{\bf (B) $V_{4}$ case}

From $V_{\omega}=0$, we obtain 
\bea
\tilde{\phi}_{\rm max}^{2}=\frac{1- \sqrt{1-2\epsilon^2}}{2} .
\label{V4-solution}
\eea
In the lower limit $\epsilon^2 \to 0$, we have $\tilde{\phi}_{\rm max}\simeq \epsilon$.
Substituting this and (\ref{Rmax-evaluate}) into (\ref{tq-evaluate}), we have
\bea\label{limitV4}
\tE\sim{1\over\epsilon^3}\frac12\tilde{\omega}^2\epsilon^2\ra\infty,&&
\tQ\sim\tilde{\omega}{1\over\epsilon^3}\epsilon^2\ra\infty,\nn
\te\sim{1\over\epsilon^2}\frac12\tilde{\omega}^2\epsilon^2\ra{\rm const.},&&
\tq\sim\tilde{\omega}{1\over\epsilon^2}\epsilon^2\ra{\rm const.},
\eea
which agree with the numerical results in Table III.
This explains why the results between Q-tubes and Q-balls are different in this model while no qualitative difference appears in the $V_3$ model.

\vspace{5mm}

{\bf (C) $V_{\rm grav.}$ case}

The solution of $V_{\omega}=0$ is 
\bea
\tilde{\phi}_{\rm max}=e^{-\frac{\epsilon^2}{2K}} .
\label{Vgrav-solution}
\eea
We note that dependence on $K$ is exteremely large. 
$\tilde{\phi}_{\rm max}$ approaches zero in the lower limit $\epsilon^2 \to -\infty$. 
Since $R_{\rm max}$ does not diverge, 
\beq
\tQ,~\tE,~\tq,~\te\ra0,
\eeq
which agree with the numerical results in Table IV.

In a realistic situation, we anticipate that $V_{\rm grav.}$ has also the nonrenormalization 
term $\tilde{V}_{\rm NR}=\beta \tilde{\phi}^{n}$ where $\beta >0$ and $n>2$. 
This does not change the qualitative behavior in the lower limit. 
However, in the upper limit, $V_{\omega}=0$ has degenerate solutions as in Type I models.
Therefore, we anticipate that the charge-energy relation for $V_{\rm grav.}$ with $\tilde{V}_{\rm NR}$ is similar to that for Type I solutions in the $V_{3}$ model.

\vspace{5mm}

{\bf (D) $V_{\rm gauge}$ case}

We should solve
\bea
\ln(1+\tilde{\phi}_{\rm max}^2)=\frac{\tilde{\omega}^{2}\tilde{\phi}_{\rm max}^2}{2} .
\label{Vgauge-solution}
\eea
In the lower limit $\epsilon^2 \to 0$, if we use Maclaurin expansion and 
neglect higher order terms $O(\tilde{\phi}_{\rm max}^5)$, we have 
\bea
\tilde{\phi}_{\rm max}^2\epsilon^2\simeq  \tilde{\phi}_{\rm max}^4.
\label{Vgauge-solution2}
\eea
Then, we obtain 
\bea
\tilde{\phi}_{\rm max}\simeq \epsilon ,
\label{Vgauge-solution3}
\eea
as in the $V_{\rm 4}$ model. Therefore, the limit values are identical to (\ref{limitV4}), which agree with the numerical results in Table V.
We also understand why the results for $V_4$ with $\tm^2>1/2$ and for $V_{\rm gauge}$ are qualitatively the same.



\section{Summary and Discussions}

We have made a comparative study of Q-balls and Q-tubes.
First, we have investigated their equilibrium solutions for four types of potentials.
The charge-energy relation depends on potential models. 
We have also noted that in some models the charge-energy relation is similar between Q-balls and Q-tubes while in other models the relation is quite different between them.
To understand what determines the charge-energy relation, which is a key of stability of 
the equilibrium solutions, we have established an analytical method to obtain the two limit values of the energy and the charge.
Our results have indicated how the existent domain of solutions and their stability depends on their shape as well as potentials.
This method would also be useful for other Q-objects or those in higher-dimensional spacetime.
These are our next subjects.

\acknowledgments
We would like to thank Kei-ichi Maeda for continuous encouragement. 
The numerical calculations were carried out on SX8 at  YITP in Kyoto University. 



\begin{thebibliography}{99}
\bibitem{Kus97b-98}
A. Kusenko, Phys.Lett. B 405, 108 (1997) 108; Nucl. Phys. B (Proc. Suppl.) 62A-C, 248 (1998).
\bibitem{AD}I. Affleck and M. Dine, Nucl. Phys. B {\bf249} 361 (1985).
\bibitem{SUSY}
K. Enqvist and J. McDonald, Phys. Lett. B {\bf 425}, 309 (1998); Nucl. Phys. B {\bf 538}, 321 (1999);
S. Kasuya and M. Kawasaki, Phys. Rev. D {\bf62}, 023512 (2000).
\bibitem{SUSY-DM}
A. Kusenko and M. Shaposhnikov, Phys. Lett. B {\bf418}, 46 (1998);
I. M. Shoemaker and A. Kusenko, Phys. Rev. D {\bf80}, 075021 (2009).
\bibitem{Kus98}A. Kusenko {\it et al.} Phys. Lett. B {\bf423} 104, (1998).
\bibitem{stability}
A. Kusenko, Phys. Lett. B {\bf404}, 285 (1997); {\bf406}, 26 (1997);
T. Multamaki and I. Vilja, Nucl. Phys. B {\bf 574}, 130 (2000);
M. Axenides, S. Komineas, L. Perivolaropoulos and M. Floratos, Phys. Rev. D {\bf 61}, 085006 (2000);
M. I. Tsumagari, E. J. Copeland, and P. M. Saffin, {\it ibid.}\ {\bf 78}, 065021 (2008).
\bibitem{PCS01}F. Paccetti Correia and M. G. Schmidt, Eur. Phys. J. {\bf C21}, 181 (2001).
\bibitem{SS}N. Sakai and M. Sasaki, Prog. of Theor. Phys., {\bf 119}, 929 (2008).
\bibitem{TS}T. Tamaki and N. Sakai, Phys. Rev. D {\bf81}, 124041 (2010); {\it ibid.}\ {\bf83}, 044027 (2011);
{\it ibid.}\ {\bf83}, 084046 (2011); {\it ibid.}\ {\bf84}, 044054 (2011).
\bibitem{strings}For a review, see, e.g., 
A. Vilenkin and E.P.S. Shellard, {\it Cosmic Strings and Other Topological Defects}, Cambridge (1994). 
\bibitem{SIN}N. Sakai, H. Ishihara and K. Nakao, Phys. Rev. D {\bf84}, 105022 (2011). 
\bibitem{EJ01}K. Enqvist and A. Jokinen, T. Multamaki, and I. Vilja,
Phys. Rev. D, {\bf63}, 083501 (2001);
E.J. Copeland and M.I. Tsumagari, {\it ibid.}\ {\bf80}, 025016 (2009);
T. Hiramastu, M. Kawasaki, and F. Takahashi, JCAP {\bf06}, 008  (2010).
\bibitem{Tsuma}R. Battye and Paul Sutcliffe, Nucl. Phys. B {\bf 590} 329 (2000); 
M.I. Tsumagari, http://www.nottingham.ac.uk/ 
$\sim$ppzphy7/webpages/people/Mitsuo/welcome.html
\bibitem{TCS}M.I. Tsumagari, E.J. Copeland, and P.M. Saffin, Phys. Rev. D {\bf78}, 065021 (2008).
\bibitem{Col85}S. Coleman, Nucl. Phys. {\bf B262}, 263 (1985). 
\bibitem{Kim}Y. Kim, K. Maeda, and N. Sakai, Nucl. Phys. {\bf B481} 453, (1996);
Y. Kim, S. J. Lee, K. Maeda, and N. Sakai, Phys. Lett. B {\bf452}, 214 (1999).
\bibitem{PS78}For a review of catastrophe theory, see, e.g., 
T. Poston and I.N. Stewart, {\it Catastrophe Theory and Its Application}, Pitman (1978). 
\end{thebibliography}
\end{document}